\documentclass[preprint,notoc]{JHEP3}
\usepackage{epsfig}

\def\eslt{\not\!\!{E_T}}
\def\eslt{E_T^{\rm miss}}

\def\to{\rightarrow}

\def\bi{\begin{itemize}}
\def\ei{\end{itemize}}
\def\te{\tilde e}

\def\tH{\tilde H}

\def\tu{\tilde u}

\def\tb{\tilde b}
\def\tf{\tilde f}
\def\td{\tilde d}

\def\tH{\tilde H}
\def\tst{\tilde t}
\def\ttau{\tilde \tau}
\def\tmu{\tilde \mu}
\def\tg{\tilde g}
\def\tnu{\tilde\nu}

\def\tq{\tilde q}

\def\tw{\widetilde W}
\def\tz{\widetilde Z}
\def\alt{\stackrel{<}{\sim}}
\def\agt{\stackrel{>}{\sim}}
\def\be{\begin{equation}}  
\def\ee{\end{equation}}  

\title{Mixed Higgsino Dark Matter\\ 
from a Large SU(2) Gaugino Mass}
\author{Howard Baer$^a$, Azar Mustafayev$^b$, Heaya Summy$^a$
and  Xerxes Tata$^c$\\
$^a$Department of Physics, Florida State University Tallahassee, 
FL 32306, USA\\
$^b$Dept. of Physics and Astronomy,
University of Kansas, Lawrence, KS 66045, USA\\
$^c$Department of Physics and Astronomy, University of Hawaii,
Honolulu, HI 96822, USA\\
E-mail: \email{baer@hep.fsu.edu},\email{amustaf@ku.edu},
\email{heaya@hep.fsu.edu},\email{tata@phys.hawaii.edu}}

\preprint{\vbox{FSU-HEP-070829, UH-511-1110-07}}

\abstract{ We observe that in SUSY models with non-universal GUT scale
gaugino mass parameters, raising the GUT scale $SU(2)$ gaugino mass
$|M_2|$ from its unified value results in a smaller value of
$-m_{H_u}^2$ at the weak scale. By the electroweak symmetry breaking
conditions, this implies a reduced value of $\mu^2$ {\it vis \`a vis}
models with gaugino mass unification.
The lightest neutralino can then be mixed Higgsino dark matter with a
relic density in agreement with the measured abundance of cold dark
matter (DM). 
We explore the phenomenology of this high $|M_2|$ DM model.
The spectrum is characterized by a very large wino mass and a
concomitantly large splitting between left- and 
right- sfermion masses. In addition, the lighter chargino and 
three light neutralinos are 
relatively light with substantial higgsino components.
The higgsino content of the LSP implies large rates for
direct detection of neutralino dark matter, and enhanced rates for its
indirect detection relative to mSUGRA.  
We find that experiments at the LHC should be able to discover SUSY over
the portion of parameter space where $m_{\tg} \alt 2350-2750$~GeV,
depending on the squark mass, while a 1~TeV electron-positron collider
has a reach comparable to that of the LHC. The dilepton mass spectrum in
multi-jet + $\ell^+\ell^- + \eslt$ events at the LHC will likely show
more than one mass edge, while its shape should provide indirect
evidence for the large higgsino content of the decaying neutralinos.
}
\keywords{Supersymmetry Phenomenology, Supersymmetric Standard Model, %
Dark Matter}

\begin{document}

\section{Introduction and framework}
\label{sec:intro}

There is overwhelming evidence showing that most of the matter in the
Universe is not baryonic, but composed of a new massive stable (or at
least very long-lived), weakly (or super-weakly) interacting particle
not contained in the Standard Model (SM) of particle physics. Moreover,
the mass density of
this cold ``dark matter'' (DM) has been precisely measured: combining the
results from the WMAP Collaboration with those from the Sloan Digital
Sky Survey gives~\cite{wmap}
\begin{equation}
\Omega_{\rm DM}h^2 = 0.111^{+0.011}_{-0.015} \ \ (2\sigma)\;,
\label{wmap}
\end{equation}
where $\Omega =\rho/\rho_c$ with $\rho_c$ the closure density of the
Universe, and $h$ is the scaled Hubble parameter, $h=0.73\pm 0.04$.
While the mass density of DM is rather precisely known, the identity of
DM remains
a mystery.  Like visible matter, the DM may well consist of
several components, in which case the density in (\ref{wmap}) yields an
{\it upper limit} on the density of any single component.
Supersymmetric (SUSY) models of particle physics with a conserved
$R$-parity include a stable, massive interacting particle, often the
lightest neutralino $\tz_1$.  Remarkably, the properties of the
neutralino are just right to enable it to serve as {\it thermally
produced} dark matter within the standard Big Bang cosmology if the SUSY
mass scale is $\sim 100$~GeV~\cite{haim,griest}.\footnote{It is possible
to invoke models where the DM results from late decays of heavy
particles and so is not in thermal equilibrium, or to invoke
non-standard Big Bang cosmologies~\cite{graciela} to obtain the observed
value of the DM density. It is, nevertheless, appealing if we can
account for the relic density data with minimal assumptions: {\it i.e.}
via thermally produced relics within the standard Big Bang model.}

A potential sticking point in the discussion of SUSY DM is that the
non-observation of direct or indirect effects of SUSY are beginning to
push $M_{\rm SUSY}$ beyond 100~GeV. As a result, the neutralino
annihilation cross-section which is $\propto 1/M_{\rm SUSY}^2$ is
correspondingly reduced, and the calculated neutralino relic abundance
is typically considerably larger~\cite{wmapcon} than the value in
Eq.~(\ref{wmap}). For instance, in the mSUGRA model~\cite{msugra}, almost
all of parameter space is excluded by the precisely
measured abundance, and only a
few distinct regions where neutralino annihilation is enhanced survive:
the nearly excluded bulk region with low masses already mentioned above
\cite{bulk}, the stau\cite{stau} or stop\cite{stop} co-annihilation
regions, the hyperbolic branch/focus point (HB/FP) region at large
$m_0$\cite{hb_fp}, where the $\tz_1$ becomes mixed higgsino dark matter
(with enhanced annihilation to $W$ and $Z$ particles via its higgsino
component), or the $A$ or $h$ resonance annihilation (Higgs funnel)
region\cite{Afunnel,drees_h}.  Within the mSUGRA framework, each of
these regions leads to characteristic patterns in collider signals at
the soon-to-be-operational CERN Large Hadron Collider (LHC), and also
to differences in signals in direct and indirect DM search experiments under
way. 

We recognize, however, that these conclusions about the expected
patterns are specific to the mSUGRA model, and can be obviated in simple
extensions of this framework where new mechanisms may arise to match the
predicted neutralino abundance to the measured CDM density. In previous
works, in the scalar sector, non-universal soft masses for the different
generations\cite{nmh}, or for the Higgs scalars have been
considered\cite{nuhm}.  In the gaugino sector, by abandoning gaugino
mass universality at the GUT scale, a variety of new mechanisms
emerge.\footnote{Non-universal gaugino masses can be accommodated in
SUSY GUT if the auxiliary field that breaks supersymmetry also breaks
the GUT symmetry \cite{nonunivgaug}.}  As shown in
Ref.~\cite{mwdm,auto}, allowing the weak scale gaugino masses $M_1\sim
M_2$ gives rise to mixed wino dark matter (MWDM), wherein $\tz_1\tz_1\to
W^+W^-$ is enhanced in the early Universe. Alternatively, if GUT scale
parameters are such that $M_1\sim -M_2$ at the weak scale, then there is
little bino-wino mixing, but neutralino
annihilation is enhanced in the early Universe because of 
bino-wino co-annihilation\cite{bwca} (BWCA).  Finally,
if the $SU(3)$ gaugino mass $M_3\ll M_1\sim M_2$ at the GUT scale, then
the Higgs mass parameter $m_{H_u}^2$ is driven to less negative values
at the weak scale, so that $\mu^2 \sim -m_{H_u}^2$ is also small,
resulting in mixed higgsino dark matter (MHDM)~\cite{belanger,m3dm}.
These models, wherein the composition of the neutralino is adjusted to
get the correct dark matter abundance, are collectively dubbed
``well-tempered neutralino'' models\cite{wtn}, with typical
neutralino-nucleon scattering
cross sections $\sigma (\tz_1 p)\sim 10^{-8}$~pb \cite{wtn_dm}, which
is within an order of magnitude of the sensitivity of current experiments.
It
is instructive to note that each of these alternatives can arise in the
top-down approach of string-inspired mixed moduli-anomaly mediated SUSY
breaking (mirage unification) models\cite{mmamsb}. It should also be
noted that the low $M_3$ framework with its concomitantly light squarks
offers a novel possibility for getting agreement with (\ref{wmap}) if
the rate for $\tz_1\tz_1 \to t\bar{t}$ is unsuppressed because $\tst_1$
is relatively light~\cite{stevem}; the phenomenology of such a scenario
has recently been detailed in Ref.~\cite{comp}.

In this paper, we present a novel possibility, again based on
non-universal gaugino mass parameters at the GUT scale, for obtaining
agreement with (\ref{wmap}) via MHDM.  We assume that the MSSM as the correct
effective field theory valid between energy scales $Q\sim M_{\rm weak}$ and
$Q = M_{GUT}\sim 2\times 10^{16}$~GeV (for text book accounts, see
Ref.~\cite{wss,spart}). We adopt the usual universality of scalar mass
parameters and trilinear scalar couplings but take GUT scale boundary
conditions in the gaugino sector of the form $M_2\gg M_1\sim M_3$. We
then inspect the evolution of the soft term 
$m_{H_u}^2$ whose 1-loop RGE is given by 
\be \frac{dm_{H_u}^2}{dt} =\frac{2}{16\pi^2}\left( -{3\over
5}g_1^2M_1^2-3g_2^2M_2^2+{3\over 10}g_1^2 S+3f_t^2X_t\right) , 
\ee 
where $t=\log Q^2$, $f_t$ is the top quark Yukawa coupling,
$X_t=m_{Q_3}^2+m_{\tst_R}^2+m_{H_u}^2+A_t^2$ and $S=m_{H_u}^2-m_{H_d}^2
+ Tr({\bf m}_Q^2- {\bf m}_L^2-2{\bf m}_U^2+{\bf m}_D^2 +{\bf m}_E^2)$.
Usually, the $f_t^2X_t$ term overcomes the upward push from the
gauge-gaugino terms (proportional to the gaugino mass parameters) and
drives $m_{H_u}^2$ to lower values as $Q$ is reduced, and ultimately to
negative values, the celebrated radiative electroweak symmetry breaking
(REWSB) mechanism\cite{rewsb}.  In the case where $M_2$ is very large at
the GUT scale, the gaugino terms initially win resulting in an {\it
upwards} push at the start of the $m_{H_u}^2$ evolution. The large $M_2$
also increases the various left- scalar soft masses to initially large
values, thus enhancing the magnitude of $X_t$ which results in an
increased {\it downward} push of the $f_t^2 X_t$ term. The resulting
value of $m_{H_u}^2({\rm weak})$ depends on the value of $M_2({\rm
GUT})$; by adjusting the latter we can arrange things so that by the
time the weak scale is reached, the (incomplete) cancellation between
the upwards and downwards push results in a negative value of
$m_{H_u}^2$ that is significantly smaller in magnitude than in models
where GUT scale gaugino mass parameters have a common value.  The weak
scale value of $\mu^2$ (at tree-level) is then obtained from the weak
scale parameters of the Higgs sector via the EWSB relation,
\be
\mu^2=\frac{m_{H_d}^2-m_{H_u}^2\tan^2\beta}{(\tan^2\beta
-1)}-{M_Z^2\over 2} .  
\ee 
We see that if $|m_{H_u}^2| \gg M_Z^2$ and for moderate to large values
of $\tan\beta$, $\mu^2 \sim -m_{H_u}^2$.  Thus, the small $|m_{H_u}^2|$
value results in a smaller $|\mu |$ parameter, and a correspondingly
larger higgsino component of the lightest neutralino $\tz_1$.  As
discussed above, MHDM can easily be compatible with the observed relic
density because the neutralino annihilation into $WW$, $ZZ$ and $Zh$
pairs is enhanced on account  of the higgsino content of $\tz_1$. Since the
large value of the wino mass underlies the root of this scenario, we
will call it the high $M_2$ dark matter (HM2DM) model.

\FIGURE[htb]{
\epsfig{file=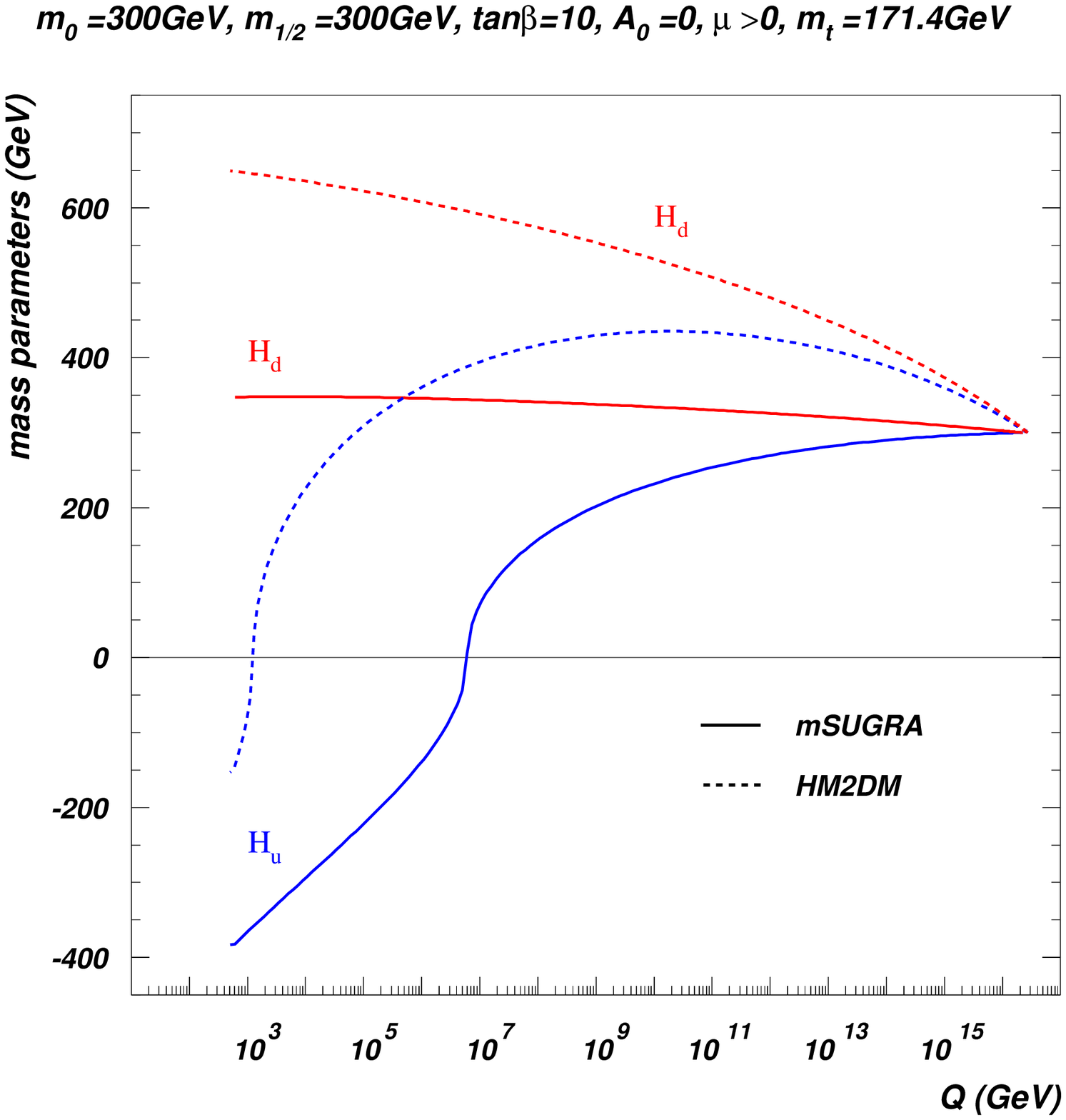,width=7cm} 
\epsfig{file=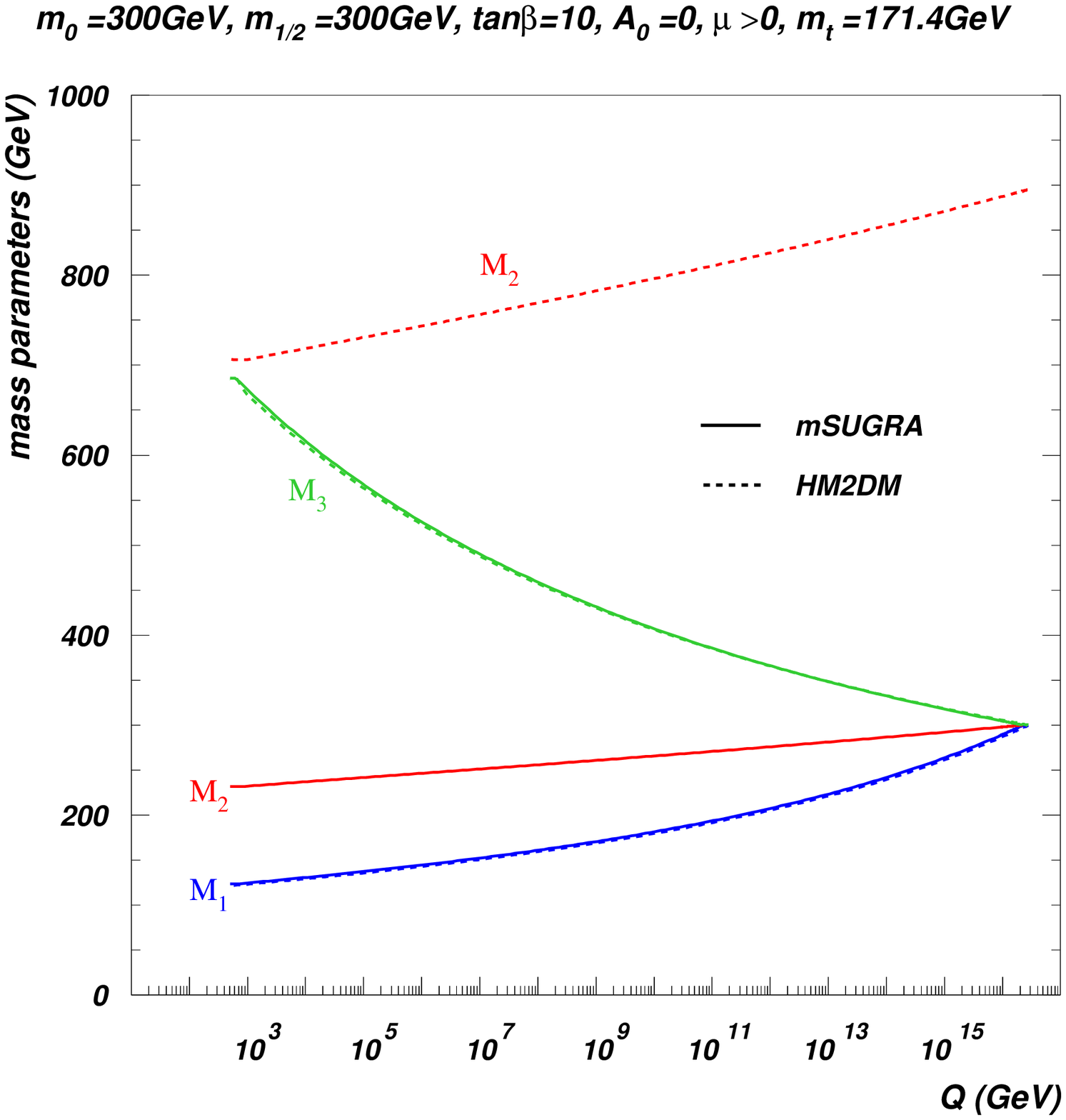,width=7cm} 
\caption{\label{fig:hi_ino} Evolution of {\it a})~the soft SUSY breaking
Higgs mass parameters ${\rm sign}(m_{H_u}^2) \sqrt{|m_{H_u}^2}|$ 
and ${\rm sign} (m_{H_d}^2) \sqrt{|m_{H_d}^2}|$, and {\it b})~the
gaugino mass parameters, as a function of the
energy scale
$Q$ in the mSUGRA model (solid) for $m_0=300$~GeV, $m_{1/2}=300$~GeV,
$A_0=0$, $\tan\beta =10$, $\mu >0$ and $m_t=171.4$~GeV, and for the
HM2DM model (dashes). The 
model parameters adopted for HM2DM are the same as in the
mSUGRA case except that $M_2=3 m_{1/2}$ at $Q=M_{GUT}$.
}}

The situation is illustrated in Fig.~\ref{fig:hi_ino}. The model is
specified by the mSUGRA parameter set augmented by one additional
parameter, $M_2$:
\be
m_0,\ m_{1/2}, \ M_2, \ A_0,\ \tan\beta\ {\rm and}\ {\rm sign}(\mu )\;,
\label{par}
\ee
where $M_1=M_3\equiv m_{1/2}$, but $M_2$ is allowed to be free (with
either sign).  We take $m_t=171.4$~GeV, in accord with recent mass
determinations\cite{mtop}.  We illustrate in Fig.~\ref{fig:hi_ino}{\it
a}) the evolution of ${\rm sign}(m_{H_u}^2) \sqrt{|m_{H_u}^2}|$ and
${\rm sign} (m_{H_d}^2) \sqrt{|m_{H_d}^2}|$ as a function of scale $Q$
in the mSUGRA model (solid) for $m_0=300$~GeV, $m_{1/2}=300$~GeV,
$A_0=0$, $\tan\beta =10$, $\mu >0$. The same running mass parameters are
shown for HM2DM for the same parameters as in the mSUGRA case except
that we now take $M_2=3 m_{1/2}$ at $Q=M_{\rm GUT}$ (dashes).  In the
mSUGRA case, $m_{H_u}^2$ evolves from a positive GUT scale value to a
large negative value at $Q=M_{\rm weak}$, resulting in a large $|\mu |$
parameter. In the case of HM2DM, however, the large value of $M_2$
 initially causes $m_{H_u}^2$ to evolve {\it upwards}, but
ultimately, the $f_t^2 X_t$ term wins out and $m_{H_u}^2$ evolves to a
not-as-large negative value and electroweak symmetry is broken.  
The smaller value of $-m_{H_u}^2$ at the weak scale leads, of course, to
a correspondingly smaller value of $|\mu |$ compared to the mSUGRA case.
In Fig.~\ref{fig:hi_ino}{\it b}), we show
the evolution of gaugino masses in mSUGRA and in the HM2DM case. For
mSUGRA, we are left with the gaugino masses at the weak scale in the
well-known ratio of $M_1:M_2:M_3\sim 1:2:7$ which implies (when $|\mu |$
is large) that $m_{\tz_1}:m_{\tw_1}:m_{\tg}\sim 1:2:7$.  In contrast, in
the HM2DM scenario, we have at the  weak scale  $M_1\ll M_3\sim M_2$,
so that we expect in general the $SU(2)$ winos to be similar in mass to
gluinos, and hence almost  decoupled at the LHC.\footnote{
The reader may well wonder whether it is possible to obtain MHDM by
increasing $M_1$ instead of $M_2$. This does not, however, appear to be
possible because the bino mass required for this 
is  so large that, though $\mu$ is indeed
reduced, the lightest neutralino dominantly becomes a wino-higgsino
mixture and annihilates too rapidly to saturate (\ref{wmap}).}

\FIGURE[tbh]{
\epsfig{file=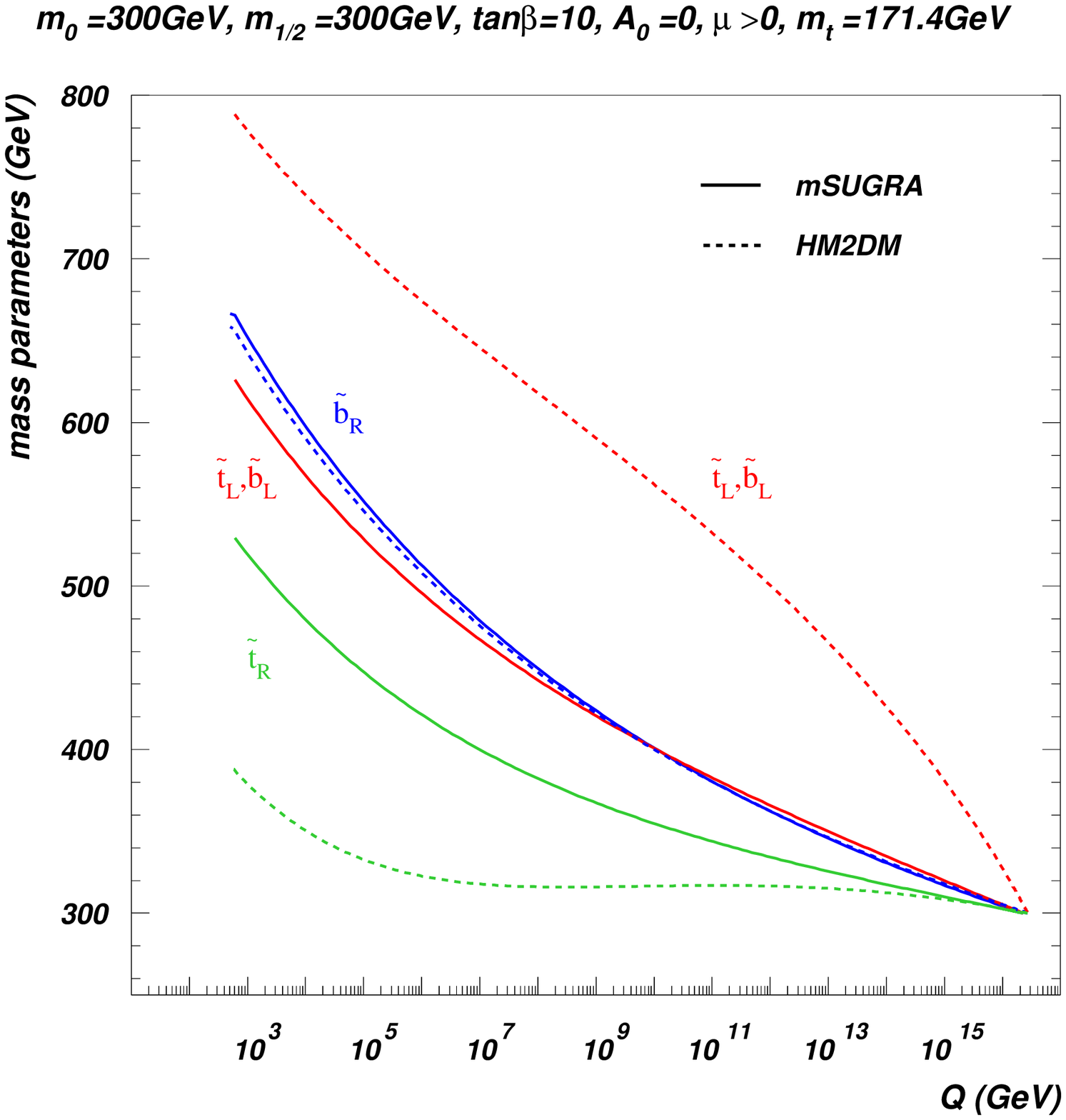,width=7cm} 
\epsfig{file=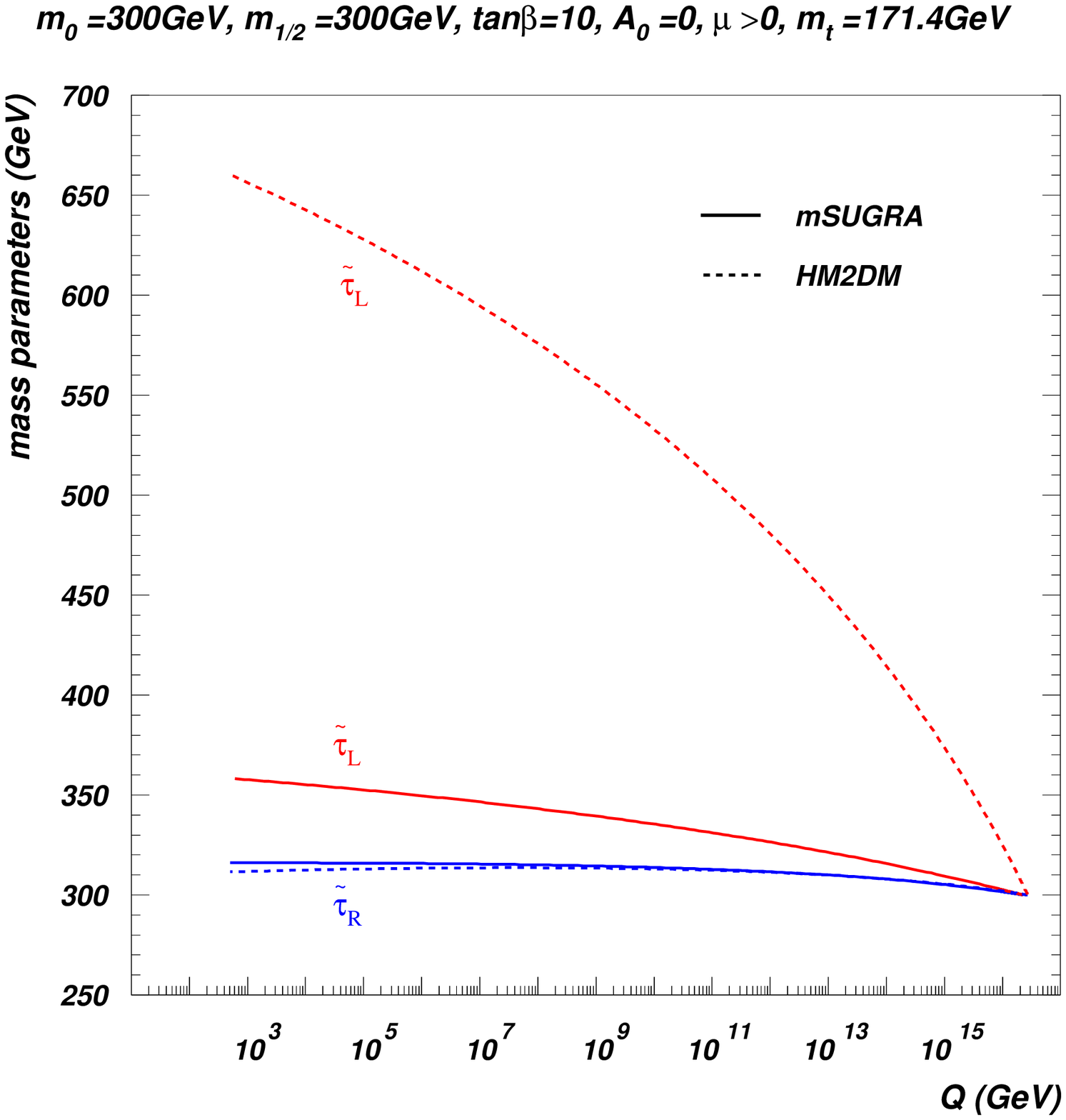,width=7cm} 
\caption{\label{fig:sq_sl} {\it a}) Evolution of the soft SUSY breaking
squark mass parameters $m_{\tst_L}$ and $m_{\tst_R}$ as a function
of scale $Q$ in the mSUGRA model (solid) for $m_0=300$~GeV,
$m_{1/2}=300$~GeV, $A_0=0$, $\tan\beta =10$, $\mu >0$ and $m_t=171.4$~GeV. The same running mass parameters are shown for HM2DM model for the same
parameters as in the mSUGRA case except that we take $M_2=3 m_{1/2}$ at
$M_{\rm GUT}$ (dashes).  {\it b}) Evolution of third generation soft SUSY
breaking slepton mass parameters $m_{\ttau_L}$ and $m_{\ttau_R}$ as
a function of scale $Q$ for the same case as in frame {\it a}).  }}

The large value of $M_2$ in the HM2DM model also influences the 
evolution of matter scalar mass parameters.  
In Fig.~\ref{fig:sq_sl}{\it a}), we show the evolution
of third generation squark mass soft parameters versus energy scale for
the mSUGRA and the HM2DM models. In the mSUGRA case, we see the
hierarchy $m_{\tb_R}>m_{\tst_L}>m_{\tst_R}$ generated by the large top
quark Yukawa coupling. However, in the HM2DM case, the large value of
$M_2$ causes left-matter scalars (which have gauge couplings to the heavy
winos) to evolve to larger values, so we
get large weak scale value of $m_{\tst_L} (=m_{\tb_L}) $, and
also for first and second generation left-sfermions. In addition,
the large $m_{\tst_L}$ enters via the  $f_t^2 X_t$ term 
in the RGE for $m_{\tst_R}^2$, causing it to reduce at the
weak scale.  Thus, in HM2DM,
we expect large $L-R$ splitting in the squark sector, with $m_{\tst_L}(=
m_{\tb_L}) >m_{\tb_R}>m_{\tst_R}$. As a result, in contrast to the
mSUGRA framework, the lighter sbottom quark $\tb_1$ is dominantly
$\tb_R$ in the HM2DM scenario.

In Fig.~\ref{fig:sq_sl}{\it b}), we show the evolution of third
generation slepton masses. In the mSUGRA case, there is only a small
intra-generation splitting at the weak scale. In the HM2DM scenario,
$m_{\ttau_R}$ evolves as in mSUGRA, but $m_{\ttau_L}$ gains a big
enhancement from the large value of $M_2$.  Thus again, in the slepton
sector, we expect large left-right splitting of slepton mass parameters
in HM2DM, with left sleptons much heavier than the right sleptons. A
similar behaviour is expected for the first two generations of sleptons.

The upshot is that if a large $M_2$ parameter is the underlying reason
for MHDM in our Universe, then characteristic sparticle mass spectra and
sparticle mixing patterns should emerge as a result.  Our goal in this
paper is to lay out the phenomenology of this framework.  In
Sec.~\ref{sec:pspace} we explore the sparticle mass and mixing patterns,
and delineate the allowed parameter space in the HM2DM scenario. In
Sec.~\ref{sec:bsg}, we examine the low energy constraints from $b\to
s\gamma$ and the SUSY contributions to the anomalous magnetic moment of
the muon. In Sec.~\ref{sec:dm}, we examine prospects for direct and
indirect dark matter detection in the HM2DM framework. In
Sec.~\ref{sec:col}, we examine collider implications of the
HM2DM scenario.  We end in Sec.~\ref{sec:conclude} with a summary of our
results.

\section{Parameter space, relic density  and mass spectrum}
\label{sec:pspace}

As discussed in the last section, the HM2DM model is completely
specified by the 
parameter set (\ref{par}),
$$
m_0,\ m_{1/2},\ M_2,\ A_0,\ \tan\beta \ {\rm and}\ {\rm sign}(\mu ),
$$
where we assume that
$M_1=M_3\equiv m_{1/2}\ge 0$ at $Q=M_{GUT}$, and where $M_2$ can assume
either sign.  The assumed equality of $M_1$ and $M_3$ can be relaxed
somewhat and our conclusions suffer little qualitative change so long as
$M_2\gg M_1$.  To calculate the sparticle mass spectrum, we use
Isajet 7.76\cite{isajet}, which allows for the input of non-universal
scalar and gaugino masses in gravity mediated SUSY breaking models where
electroweak symmetry is broken radiatively. The relic density is
evaluated via the IsaReD program\cite{isared}, which is part of the
Isatools package.  IsaReD evaluates all $2\to 2$ tree level neutralino
annihilation and co-annihilation processes and implements relativistic
thermal averaging in the relic density calculation.

In the upper frame of Fig.~\ref{fig:rd}, we show the neutralino relic
density for the parameter space point $m_0=m_{1/2}=300$~GeV, $A_0=0$,
$\tan\beta =10$ and $\mu >0$, versus the ratio $r_2=M_2({\rm
GUT})/m_{1/2}$. For $r_2=1$ corresponding to the mSUGRA model,
$\Omega_{\tz_1}h^2=1.1$, so that the point is strongly excluded because
it yields too much dark matter. For $r_2\sim 0.6$, we arrive at the MWDM
case explored in Ref.~\cite{mwdm}, while for $M_2\sim -0.5$, we have
BWCA, explored in Ref.~\cite{bwca}. If we instead {\it increase} the
magnitude of $M_2$, then we find at $r_2\sim 3$ a match to the
measured relic abundance, with $\Omega_{\tz_1}h^2\sim 0.1$. This is the
case of HM2DM. As we increase $r_2$ further, the relic density starts to
go back up. This is because the neutralino mass falls below $M_Z$ and
ultimately $M_W$, so that (except neutralinos with high thermal energy)
the processes $\tz_1\tz_1 \to ZZ, \ WW$ become disallowed, and the total
annihilation cross section is correspondingly reduced. Ultimately, the
relic density again starts to drop because of the annihilation via
off-shell $Z$. 
We note here that since the gaugino masses enter the soft
SUSY breaking Higgs masses as their square, there is also a point
with good relic density at $r_2\sim -2.5$. Again, the shoulder in the
relic density curve marks where the annihilation to vector boson pairs
becomes disallowed. 

\FIGURE[tbh]{
\epsfig{file=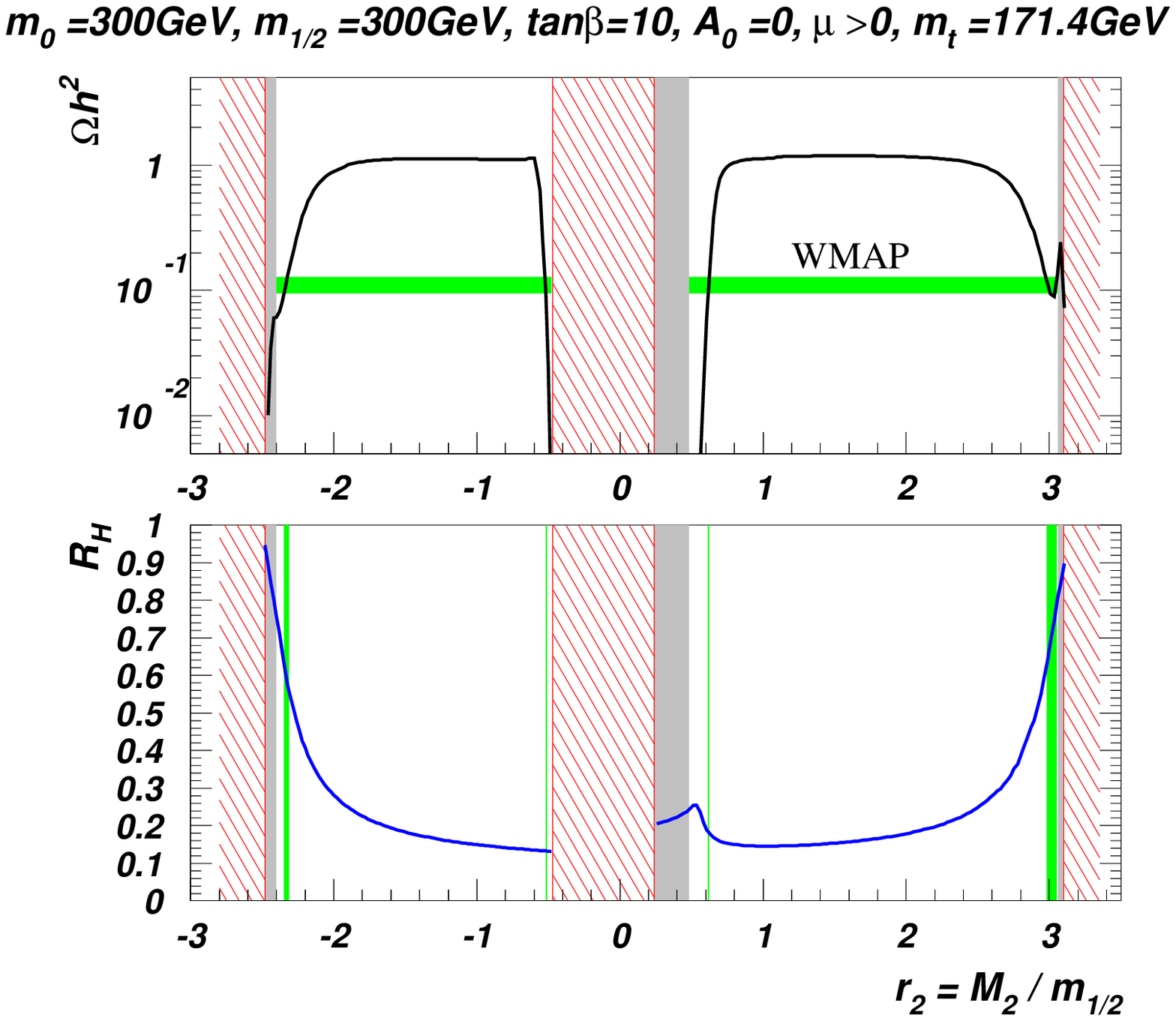,width=12cm,angle=0} 
\caption{\label{fig:rd} The neutralino relic density $\Omega_{CDM}h^2$
(upper frame) and the higgsino component $R_{\tH}$ of the lightest
neutralino (lower frame) as a function of $r_2$, for $m_0=300$~GeV,
$m_{1/2}=300$~GeV, $A_0=0$, $\tan\beta =10$, $\mu >0$ and $m_t=171.4$~GeV. In the grey regions, the chargino mass falls below its lower limit
of 103.5~GeV from LEP2 experiments, while the red hatched regions are
excluded either because the $Z$ width becomes too large, or because
electroweak symmetry is not correctly broken. The green region is
where the relic density falls in the range (\ref{wmap}). }}

We also show the higgsino content of $\tz_1$, defined by $R_{\tH}
\equiv\sqrt{v_1^{(1)2}+v_2^{(1)2}}$ (in the notation of Ref.~\cite{wss})
in the lower frame of Fig.~\ref{fig:rd}. We see that over the bulk of
the range of $r_2$, the higgsino composition of $\tz_1$ is quite low,
since $\tz_1$ is dominantly bino-like, or for $0\leq r_2 \leq 0.6$, a mixture
of wino and bino. In the case of HM2DM with $r_2\simeq 3$ or $-2.5$,
$R_{\tH}$ has risen to $\sim 0.6$, indicating  mixed
higgsino-bino dark matter.

In Fig.~\ref{fig:sigv}, we show the 
thermally averaged 
neutralino
annihilation cross section times relative velocity, integrated from
temperature $x\equiv T/m_{\tz_1}=0$ to freeze-out $x=x_F$, versus $r_2$
for the same parameter choices as in Fig.~\ref{fig:rd}.  The inverse of
this quantity enters the relic density calculation, so that a large
integrated annihilation rate leads to a small relic density. For
clarity, we display only positive values of $r_2$. We see that while
neutralino annihilation to lepton pairs via slepton exchange is dominant
in the case of mSUGRA, when we move to the case of HM2DM, where the
$\tz_1$ is a mixed bino-higgsino state, then annihilation to $WW$, $ZZ$
and $Zh$ dominates, as is typical for mixed higgsino dark matter.  As
just discussed, these cross sections drop-off near the upper end of the
range of $r_2$ once $m_{\tz_1}$ falls below $M_Z$ or $M_W$.  In this
range, annihilation via $s$-channel $Z$ (which has large couplings to
the $\tz_1$ pair on account of the large higgsino content of $\tz_1$)
dominates, and ultimately becomes resonant so that the relic density
drops below its observed value.
\FIGURE[tbh]{
\epsfig{file=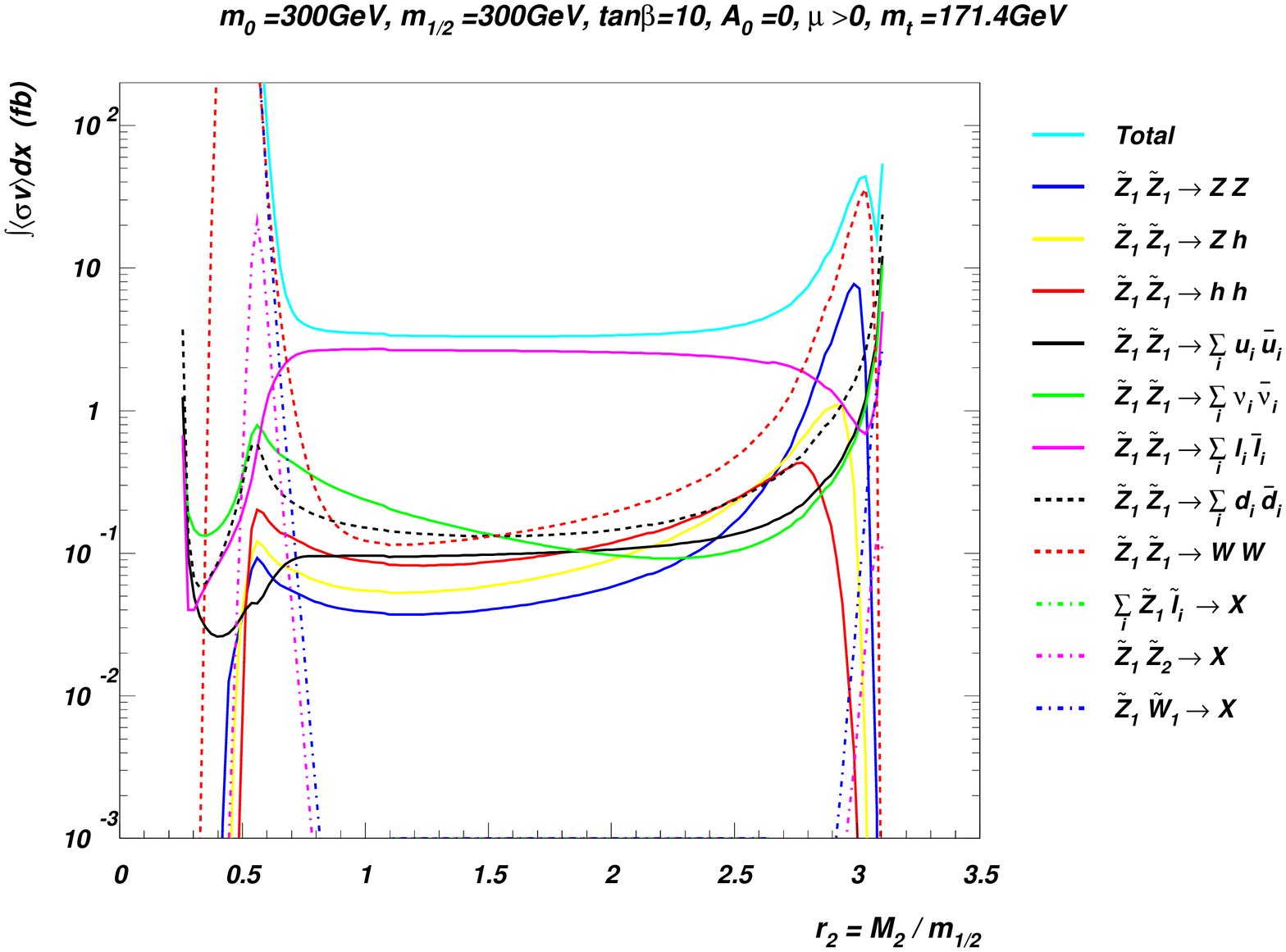,width=14cm}
\caption{\label{fig:sigv} Thermally
 averaged
neutralino annihilation
cross sections (calculated in the HM2DM model) times neutralino relative
velocity, integrated from $x=0$ to $x_F$ versus $r_2$, for the same
parameters as in Fig.~\ref{fig:mass}.}}

In Fig.~\ref{fig:mass}, we show the sparticle mass spectrum versus $r_2$
for the same parameter choice as in Fig. \ref{fig:rd}. At $r_2=1$, we
see a large mass gap $m_{\tz_2}-m_{\tz_1}\sim 100$~GeV in the case of
the mSUGRA model. As $M_2$ increases, the $\mu$ parameter decreases, and
falls rapidly beyond $r_2\sim 2$.  In the region of $r_2\sim 2.5-3$, the
curves for $m_{\tz_2}$, $m_{\tz_3}$ and $m_{\tw_1}$, and for very large
$r_2$, $m_{\tz_1}$ in place of $m_{\tz_3}$, track the $\mu$ value,
indicating that these particles are dominantly higgsino-like. We also
see that $m_{\te_L}$, and
\FIGURE[tbh]{
\epsfig{file=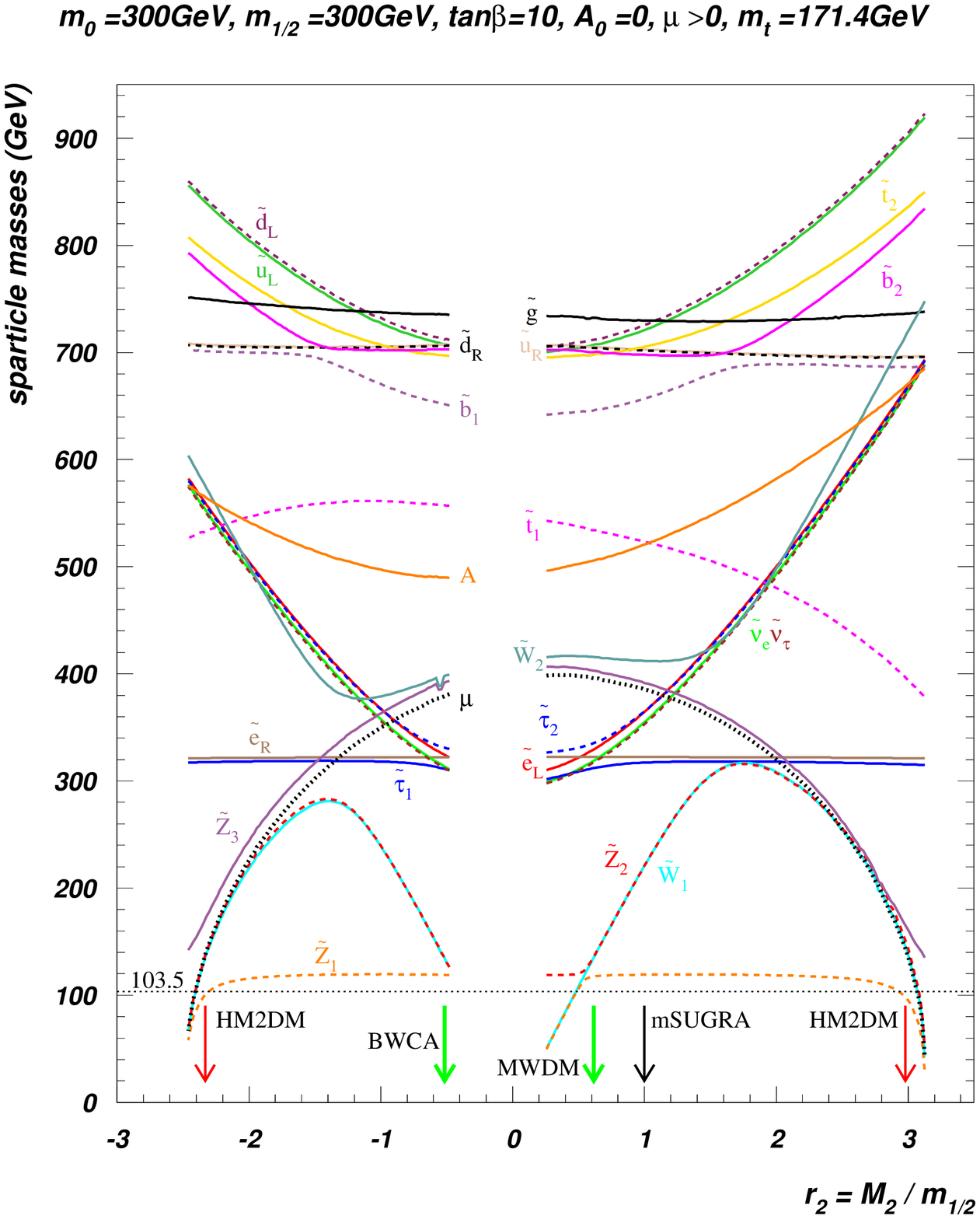,width=10cm} 
\caption{\label{fig:mass} Various sparticle and Higgs boson masses and
the $\mu$ parameter in the HM2DM model {\it versus} $r_2$ for $m_0=300$~GeV,
 $m_{1/2}=300$~GeV, $A_0=0$, $\tan\beta =10$ and $\mu >0$.}}
$m_{\tu_L,\td_L}$ all increase with increasing $r_2$, owing to the
upward push in their respective RGEs.  The value of $m_A$, given at
tree-level by $m_A^2\sim m_{H_d}^2-m_{H_u}^2$, also increases with $r_2$
as can be seen from Fig.~\ref{fig:hi_ino}{\it a}.  The right- squark and
slepton masses, on the other hand, are roughly independent of $r_2$, so
that in HM2DM we expect a larger mass gap between L and R squarks and
sleptons relative to the mSUGRA model. This also leads to the
level-crossing in the $b$-squark system that we mentioned earlier: for
small values of $r_2$ (including in the mSUGRA model) $\tb_1$ is
dominantly $\tb_L$, while for $r_2 \agt 2$, $\tb_1$ becomes mostly
$\tb_R$.  The value of $m_{\tst_1}$ actually decreases with increasing
$M_2$, which is due in part to the diminishing value of $m_{\tst_R}^2$
as shown in Fig.~\ref{fig:sq_sl}{\it a}), and in part due to an
increasingly negative weak scale value of $A_t$.  For the most part, the
figure is nearly symmetric between positive and negative values of $M_2$
since, as mentioned above, the scalar SSB RGEs contain $M_2^2$, and not
$M_2$. The $A_{t,b,\tau}$ parameter RGEs all have $M_2$ entering
linearly, so that $A$ term evolution is {\it not} symmetric between
positive and negative $\mu$. This gives rise to the unsymmetrical
behavior, most noticeable in the $m_{\tst_1}$ curve. Of course,
$m_{\tw_2}$-- which becomes essentially $|M_2|$ in the HM2DM model-- is
also asymmetric since the value of $|r_2|$ required to saturate the
relic density is itself asymmetric between positive and negative
masses. This asymmetry could have an impact upon the accessibility of
$\tw_2$ and $\tz_4$ at future electron-positron colliders.

We have already discussed how the intra-generation mixing patterns are
affected by the large value of $|M_2|$: 
since a large value of $m_{\tf_L}^2$ results when $|M_2|$ is large,  
the lighter sfermions are dominantly $\tf_R$ within the HM2DM framework.
This is confirmed in Fig.~\ref{fig:mix},
where we show the dependence of the sfermion mixing angle $\theta_f$
defined in Ref.~\cite{wss} on $r_2$
for the same parameters as in Fig.~\ref{fig:mass}. We see that the
$\tb_1$ and $\ttau_1$ are essentially $\tb_R$ and $\ttau_R$ in the HM2DM
model.

\FIGURE[tbh]{
\epsfig{file=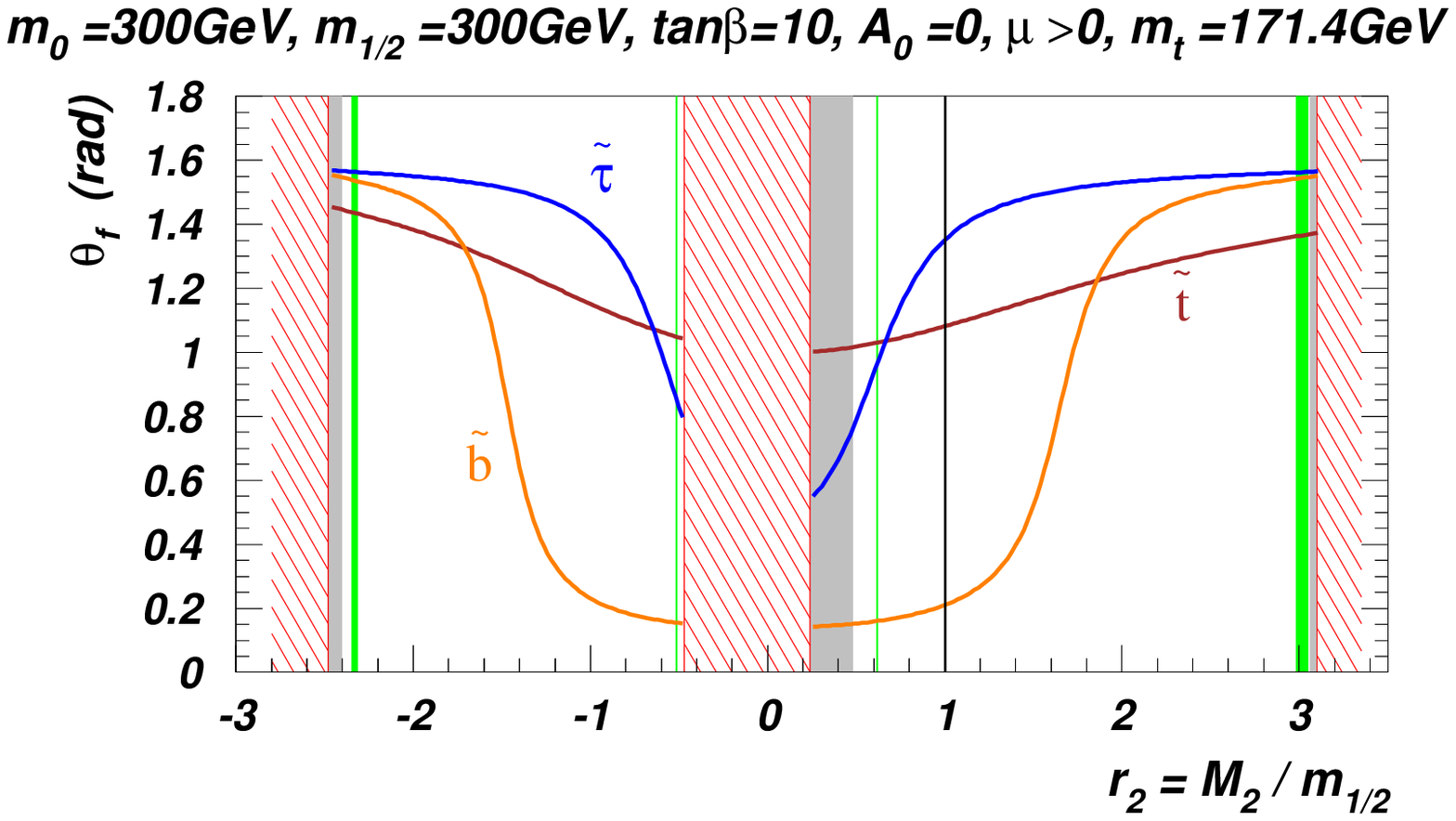,width=10cm} 
\caption{\label{fig:mix} The sfermion mixing angle $\theta_f$ as defined
in Ref.\cite{wss} for $f=t, \ b$ and $\tau$
 {\it versus} $r_2$ for $m_0=300$~GeV, 
 $m_{1/2}=300$~GeV, $A_0=0$, $\tan\beta =10$ and $\mu >0$.  We see that
for the HM2DM model, the $\theta_f$ are all large
so that the lighter states are all dominantly right sfermions.
The various shadings are the same as in Fig.~\ref{fig:rd}.
}}
\TABLE{
\begin{tabular}{lcccc}
\hline
parameter & mSUGRA & HM2DM1 & HM2DM2 & HM2DM3 \\
\hline
$m_0$ & 300 & 300 & 300 & 300 \\
$M_1$ & 300 & 300 & 300 & 300 \\
$M_2$ & 300 & 900 & -700 & -695 \\
$M_3$ & 300 & 300 & 300 & 300 \\ \hline
$\mu$ & 385.1 & 134.8 & 136.5 & -144.4 \\
$m_{\tg}$   & 729.7 & 736.4 & 749.7 & 749.5 \\
$m_{\tu_L}$ & 720.8 & 901.8 & 840.7 & 838.6 \\
$m_{\tst_1}$& 523.4 & 394.3 & 533.0 & 534.7 \\
$m_{\tb_1}$ & 656.8 & 686.4 & 701.2 & 700.7 \\
$m_{\te_L}$ & 364.5 & 669.3 & 559.9 & 557.0 \\
$m_{\te_R}$ & 322.3 & 321.3 & 321.4 & 321.4 \\
$m_{\tw_2}$ & 411.7 & 719.7 & 575.7 & 575.1 \\
$m_{\tw_1}$ & 220.7 & 136.5 & 133.9 & 144.4 \\
$m_{\tz_4}$ & 412.5 & 723.1 & 583.2 & 580.7 \\
$m_{\tz_3}$ & 391.3 & 160.2 & 170.1 & 168.4 \\ 
$m_{\tz_2}$ & 220.6 & 142.3 & 136.2 & 141.9 \\ 
$m_{\tz_1}$ & 119.2 & 94.8  & 99.9  & 108.6 \\ 
$m_A$       & 520.3 & 670.7 & 565.0 & 563.3 \\
$m_{H^+}$   & 529.8 & 679.8 & 574.3 & 572.7 \\
$m_h$       & 110.1 & 111.9 & 107.6 & 107.5 \\ \hline
$\Omega_{\tz_1}h^2$& 1.1 & 0.10 & 0.11 & 0.12 \\
$BF(b\to s\gamma)$ & $3.0\times 10^{-4}$ & $2.3\times 10^{-4}$ &
$3.3\times 10^{-4}$ & $4.0\times 10^{-4}$ \\
$\Delta a_\mu    $ & $12.1 \times  10^{-10}$ & $3.1 \times  10^{-10}$ & 
$-7.4\times 10^{-10}$ & $7.0\times 10^{-10}$ \\ 
$\sigma_{SI} (\tz_1p )$ & 
$2.1\times 10^{-9}\ {\rm pb}$ & $3.4\times 10^{-8}\ {\rm pb}$ & 
$2.5\times 10^{-8}\ {\rm pb}$ & $3.2\times 10^{-9}\ {\rm pb}$\\
$|v_1^{(1)}|$ & 0.05 & 0.40 & 0.37 & 0.32 \\
\hline
\end{tabular}
\caption{Input parameters and resultant sparticle masses in GeV units
together with the predicted neutralino relic density, direct LSP detection
scattering cross section from a proton, $B(b\to s\gamma)$ and $\Delta
a_{\mu}$, the SUSY contribution to the anomalous magnetic moment of the muon,
for mSUGRA and three HM2DM scenarios. In each case, we fix
$A_0=0$, $\tan\beta =10$ and $m_t=171.4$~GeV.
}
\label{tab:m2dm}}
%

In Table~\ref{tab:m2dm}, we list various sparticle masses and $\mu$, along with
expectations for the relic density, $a_\mu^{SUSY}$, $BF(b\to s\gamma )$
and spin-independent direct DM detection cross section
$\sigma_{SI}(\tz_1 p)$ for several different cases with all parameters
other than $M_2({\rm GUT})$ set as in Fig.~\ref{fig:mass}. The first
case, mSUGRA, provides a benchmark for comparison with the HM2DM cases.
For case HM2DM1, we take $M_2=900$~GeV, to obtain $\Omega_{\tz_1}h^2\sim
0.1$ as required by observation. As expected the gluino mass hardly
changes in going from mSUGRA to HM2DM1, while the light charginos and
neutralinos have all gotten much lighter in accord with the decreasing
$\mu$ parameter. The light chargino also changes its character from
being dominantly wino-like to dominantly higgsino-like. In fact, the
mass gap $m_{\tz_2}-m_{\tz_1}$, which was of order 100~GeV in mSUGRA,
has dropped to $\sim 50$~GeV in HM2DM1. This means the spoiler decay
modes $\tz_2\to\tz_1 h$ and $\tz_2\to\tz_1 Z$ will be closed for HM2DM,
and a dilepton mass edge should be visible in collider events where $\tz_2$ is produced at large rates either directly or via gluino and squark cascade decays.
The neutralino $\tz_3$ also
has a large higgsino component and cannot be split very much from
$\tz_2$; its leptonic decays, therefore, should also lead to a distinct
mass edge in the dilepton mass spectrum.
We also note that the left-
squarks and sleptons have become 200-300~GeV heavier than in mSUGRA,
while the masses of right- squarks and sleptons, as expected in the
HM2DM scenario, are essentially unchanged. The top squark $\tst_1$ has
become significantly lighter in the HM2DM1 case, which leads to a
deviation of the branching fraction $B(b\to s\gamma )$ from its SM
value.  In contrast, the SUSY contribution to $\Delta a_{\mu}$ is
diminished, owing to the increased left smuon and sneutrino masses. The
DM direct detection rate has increased to the $\sim 10^{-8}$~pb level
expected in models with MHDM~\cite{wtn_dm}.
In case HM2DM2, we take $M_2=-700$~GeV, which also gives the correct
relic abundance. In this case, the $\tst_1$ is heavier than in case
HM2DM1, so that $BF(b\to s\gamma )$ is more closely in agreement with
its measured value: $BF(b\to s\gamma )=(3.55\pm 0.26)\times 10^{-4}$
from a combination of CLEO, Belle and BABAR data\cite{bsg_ex}.  However,
the value of $\Delta a_{\mu}^{SUSY}$, which is proportional to
$\frac{m_\mu^2\mu M_2\tan\beta}{M_{SUSY}^4}$, has turned negative, in
contrast to the measured deviation, which is positive.  This can be
rectified by choosing in addition to $M_2<0$, $\mu <0$, as in the HM2DM3
case in the last column of the table. This case now gives a positive
contribution to $\Delta a_\mu$, but also a deviation in $BF(b\to s
\gamma )$ which is now somewhat larger than the measured value. This
latter case with opposite signs of $\mu$ and $M_1$ also gives a
significantly lower direct DM detection cross section, due to negative
interference between $h$- and $H$-mediated scattering
amplitudes~\cite{ellis,bbko}.

While our discussion up to now has been confined to  particular values
of $m_0, \ m_{1/2}, \cdots$, it should be clear that the method of
raising $|M_2|$ to obtain MHDM in agreement with (\ref{wmap}) is quite
general, although a different value of $|r_2|$ will be needed for each
point in $m_0\ vs.\ m_{1/2}$ parameter space.  We have scanned over the
$m_0\ vs.\ m_{1/2}$ parameter space for $A_0=0$ and $\tan\beta =10$ to
extract the particular $r_2$ value needed at each point to obtain the
WMAP measured CDM density.  The results are shown as contours of $r_2$
in Fig.~\ref{fig:r2} for {\it a})~$M_2 > 0$ with $\mu >0$, and {\it
b})~$M_2< 0$ with $\mu <0$.  The red-shaded regions are theoretically
excluded, due to lack of EWSB, due to a stau or (if $r_2$ is very small)
chargino LSP, or because the $Z$-width constraint is violated. The blue
shaded regions are excluded because $m_{\tw_1} < 103.5$~GeV, in
contradiction with sparticle search limits from LEP2. The green shaded
regions give $\Omega_{\tz_1}h^2<0.13$ in the mSUGRA model, so in these
regions there is no need to dial $M_2$ to large values. In frame {\it
a}), the contours in $r_2$ range from $r_2\sim 3$ in the low $m_0$,
$m_{1/2}$ region, to $r_2\alt 2$ when nearing the HB/FP region (which
already has $|\mu |$ suppression due to a large $m_0$ value).  We note
the appearance of a white region to the left of the mSUGRA stau
co-annihilation region that is allowed in the HM2DM framework; in this
case, since $|\mu |$ is reduced, the value of $m_{\tz_1}\sim |\mu |$ and
falls below $m_{\ttau_1}$. There is also a region just below
$m_{1/2}\sim 0.3$~TeV
 which turns out to be LEP2 excluded, while even smaller values
of $m_{1/2}\sim 0.2$~TeV re-emerge as LEP2 allowed.  The LEP2 excluded
region at $m_{1/2}\sim 225-290$~GeV occurs because for $m_{1/2}<290$~GeV, 
$m_{\tz_1}$ drops first below $M_Z$, then below $M_W$.  This shuts
off the $\tz_1\tz_1\to ZZ,\ WW$ annihilation modes, so that even larger
$M_2$ values are needed to drive $|\mu |$ to even smaller values.  The
lower $|\mu |$ values then drive $m_{\tw_1}$ below the LEP2 search limit
that requires $m_{\tw_1}>103.5$~GeV.  For even lower $m_{1/2}$ values,
the allowed region opens up again because, for the smaller value of
$m_{1/2}$, the higgsino components of $\tz_1$ allow for efficient
annihilation through the {\it off-shell} $s$-channel $Z$ exchange so
$|\mu |$ need not be as small, and the chargino mass can be above the
LEP2 bound in the range of $m_{1/2}\sim$150-225~GeV.
\FIGURE[tbh]{
\mbox{\epsfig{file=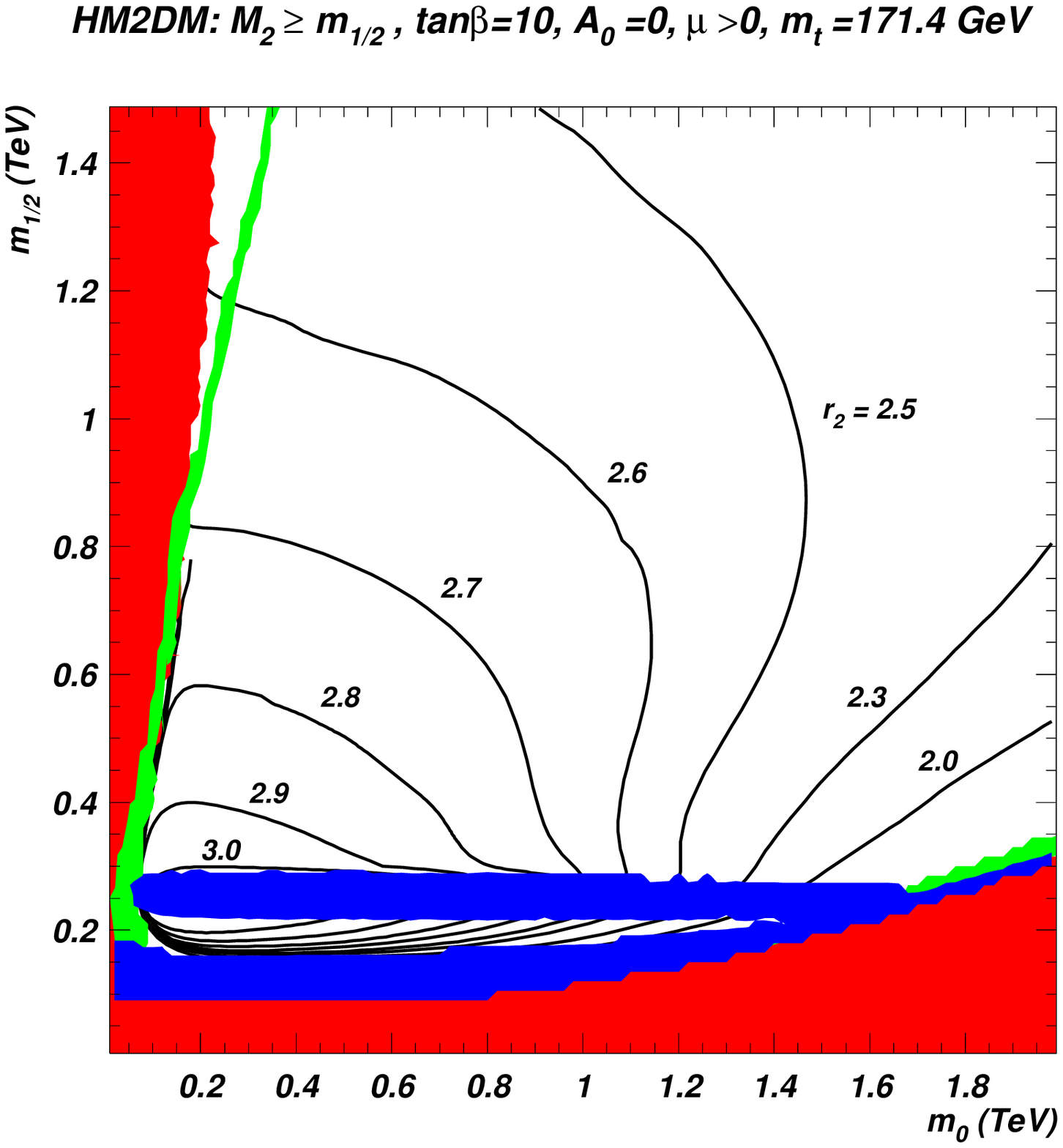,width=7cm}
\epsfig{file=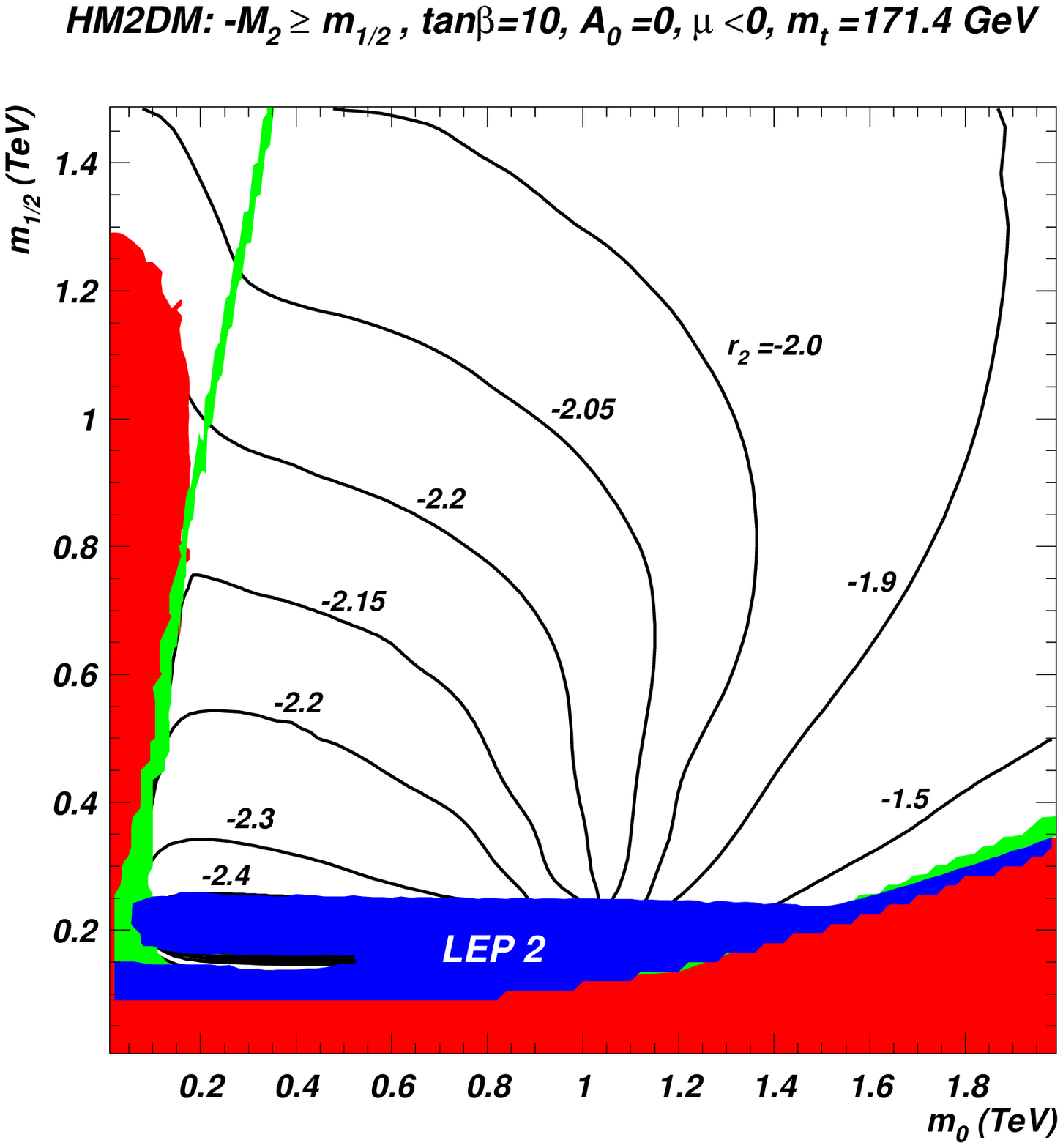,width=7cm}}
\caption{\label{fig:r2}
Contours of $r_2$ in the 
$m_0\ vs.\ m_{1/2}$ plane with
$\tan\beta =10$, $A_0=0$ for {\it a})~$M_2>0$ with  $\mu >0$, and 
{\it b})~$M_2<0$ with  $\mu <0$. Each point in these planes has
$r_2$ dialed to a high value such a value that $\Omega_{\tz_1}h^2 \simeq
0.11$. 
The red and blue regions are excluded for reasons explained in the text.
}}

The corresponding contours for negative $M_2$ are shown in
Fig.~\ref{fig:r2}{\it b}), where we also take $\mu <0$ to realize a
positive value of $\Delta a_\mu^{\rm SUSY}$; the frame  mainly differs from the
positive $M_2$ case in that the range of $|r_2|$ is somewhat lower than in
frame {\it a}), presumably because of differences in $m_{\tz_1}$ and in
the coupling of $\tz_1$ to $W$'s and $Z$'s. The LEP2-allowed region at
$m_{1/2}\sim 0.2$~TeV from frame {\it a}) now also disappears (despite
the smaller value of $r_2$,  $\mu$ is now slightly smaller than in frame
{\it a}, making it more difficult to evade the LEP 2 bound),
except for the thin sliver where $2m_{\tz_1}\sim m_h$ where, because of
the resonance enhancement, a raised value of $M_2$ is not needed to
saturate the measured relic density.

Up to now, we have confined our discussion to a fixed value of
$\tan\beta=10$. Our considerations also apply for other values of
$\tan\beta$. Specifically, we have checked that for
$m_0=m_{1/2}=300$~GeV, $A_0=0$ and $\mu >0$, that the relic density
measurement is saturated for $r_2:2.8-3.4$ for $\tan\beta \alt 40$; for
yet larger values of $\tan\beta$, the required value of $r_2$ drops
rapidly, and consistency with the upper bound on the relic density is
possible in the mSUGRA model once $\tan\beta \agt 46$ because neutralino
annihilation via $s$-channel $A$ becomes large, while for the largest
values of $\tan\beta$ co-annihilation via staus becomes dominant.

\section{${\bf b\to s\gamma}$ and ${\bf (g-2)_\mu}$}
\label{sec:bsg}

Now that we have established that any point in $m_0\ vs.\ m_{1/2}$ space
can be made dark-matter consistent by increasing $|M_2|$, we delineate
regions of parameter space where the recent measurements of the
branching fraction $BF(b\to s\gamma )$~\cite{bsg_ex} or of $(g-2)_\mu$~\cite{pdb,davier} are consistent with predictions of the HM2DM model.

\subsection{$BF(b\to s\gamma )$}

The branching fraction $BF(b\to s\gamma )$ is extremely interesting
largely because amplitudes for supersymmetric contributions mediated by 
$\tw_i\tst_j$ and $bH^+$ loops are expected to be of similar size as
the leading SM amplitude mediated by a $tW$ loop~\cite{bsg}. 
The measured branching fraction, from a combination of CLEO, Belle and
BABAR experiments~\cite{bsg_ex}, 
is $BF(b\to s\gamma )=(3.55\pm 0.26)\times 10^{-4}$, while the
latest SM calculations find\cite{bsg_th} $BF(b\to s\gamma )=
(3.29\pm 0.33)\times 10^{-4}$. In view of the good agreement between
the SM and experiment, any SUSY contribution to $BF(b\to s\gamma )$ 
should be somewhat suppressed, unless there are cancellations 
between different SUSY loops, or the summed SUSY contribution 
fortuitously turns out to be twice the SM amplitude but with
the opposite sign.

We evaluate $BF(b\to s\gamma )$ using the IsaBSG code~\cite{bsg}, a part
of the Isatools package.
The dependence of the branching fraction on $r_2$ is illustrated in
Fig.~\ref{fig:bsg1} for our canonical point 1 from Table~\ref{tab:m2dm}.
We see that in the mSUGRA
case, $BF(b\to s\gamma )$ is not far below its measured value.
Dialing $r_2$ to high positive values decreases the branching fraction
(recall that $\tst_1$ becomes lighter)
and the discrepancy with experiment grows until we hit the
green DM allowed region, where we find 
$BF(b\to s\gamma )\sim 2.3\times 10^{-4}$.
For negative values of $r_2$, $BF(b\to s\gamma )$ varies much less, and 
in the WMAP-favoured range is actually in
close accord with its measured value. For the
favoured negative sign of $\mu$, the branching fraction becomes a bit
too large when $r_2< 0$. 
\FIGURE[tbh]{
\mbox{\hspace{-1cm}
\epsfig{file=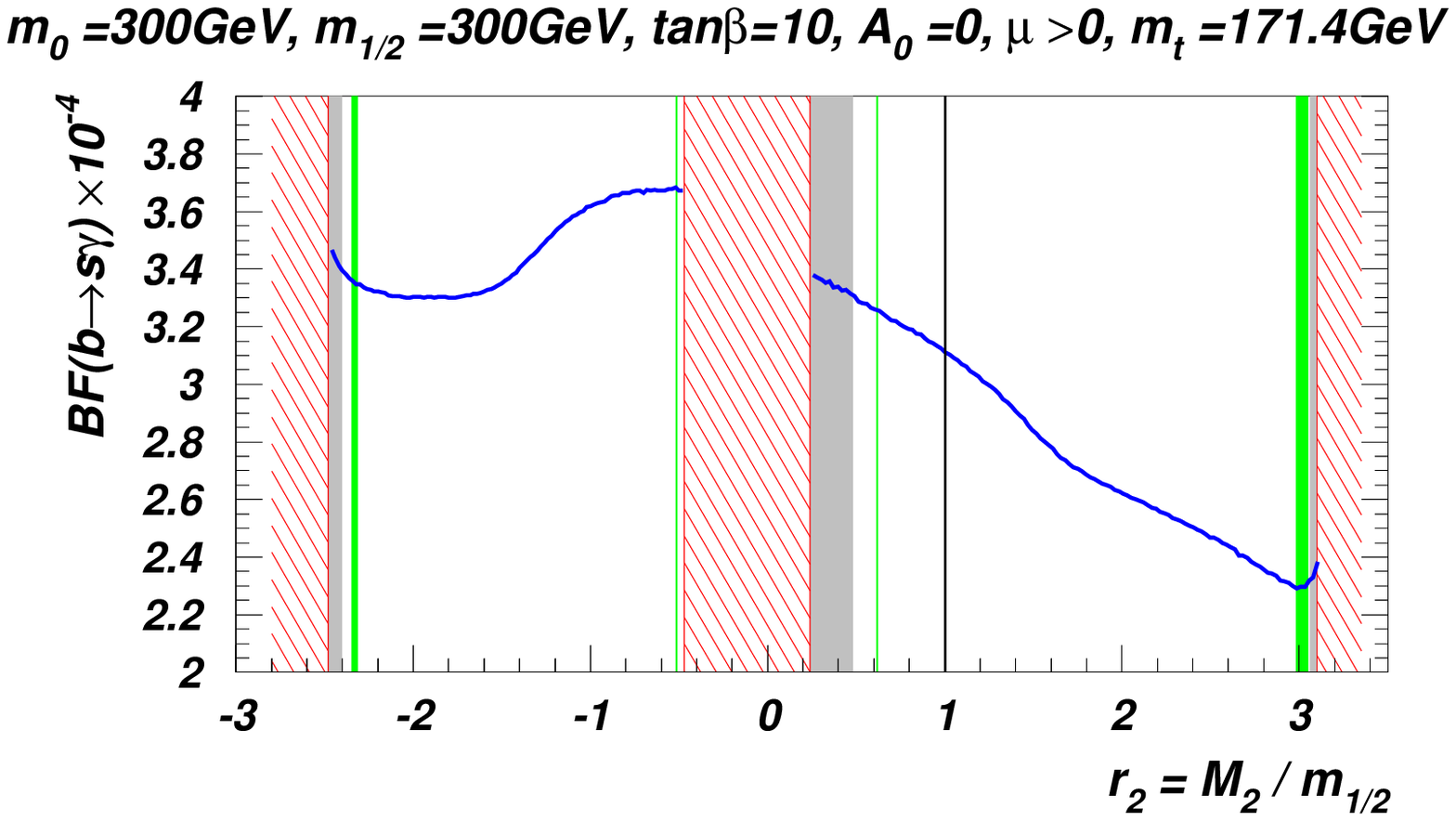,width=12cm,angle=0} }
\caption{\label{fig:bsg1}
Branching fraction $BF(b\to s\gamma )$ {\it versus} $M_2$ for
$m_0=m_{1/2}=300$~GeV, $A_0=0$, $\tan\beta =10$, and $\mu >0$.
The various shadings are the same as in Fig.~\ref{fig:rd}.
}}

While this particular point at high $M_2>0$ may seem somewhat
discouraging, in Fig.~\ref{fig:bsg2} we show by the black
contours the branching fraction  $BF(b\to
s\gamma )$ in the $m_0\ vs.\ m_{1/2}$ plane with $A_0=0$, $\tan\beta
=10$ and {\it a})~$M_2 > 0$ with 
$\mu >0$, and {\it b})~$M_2< 0$ with $\mu < 0$, 
where at each point we have dialed $|M_2|$ to high
values so that that $\Omega_{\tz_1}h^2$ saturates the measured
value in (\ref{wmap}). In frame {\it a}), we see that
$BF(b\to s\gamma )$ is low only in the very low $m_0$ and $m_{1/2}$
corner, but is not far from its measured value (considering
theoretical uncertainties) for $m_0,\ m_{1/2}\agt 0.5$~TeV.
In contrast, in frame {\it b}), the measured value of $BF(b\to s\gamma)$
clearly disfavours small values of $m_0$ and $m_{1/2}$, requiring these
to be $\agt 700$~GeV.  

\FIGURE[tbh]{
\mbox{
\epsfig{file=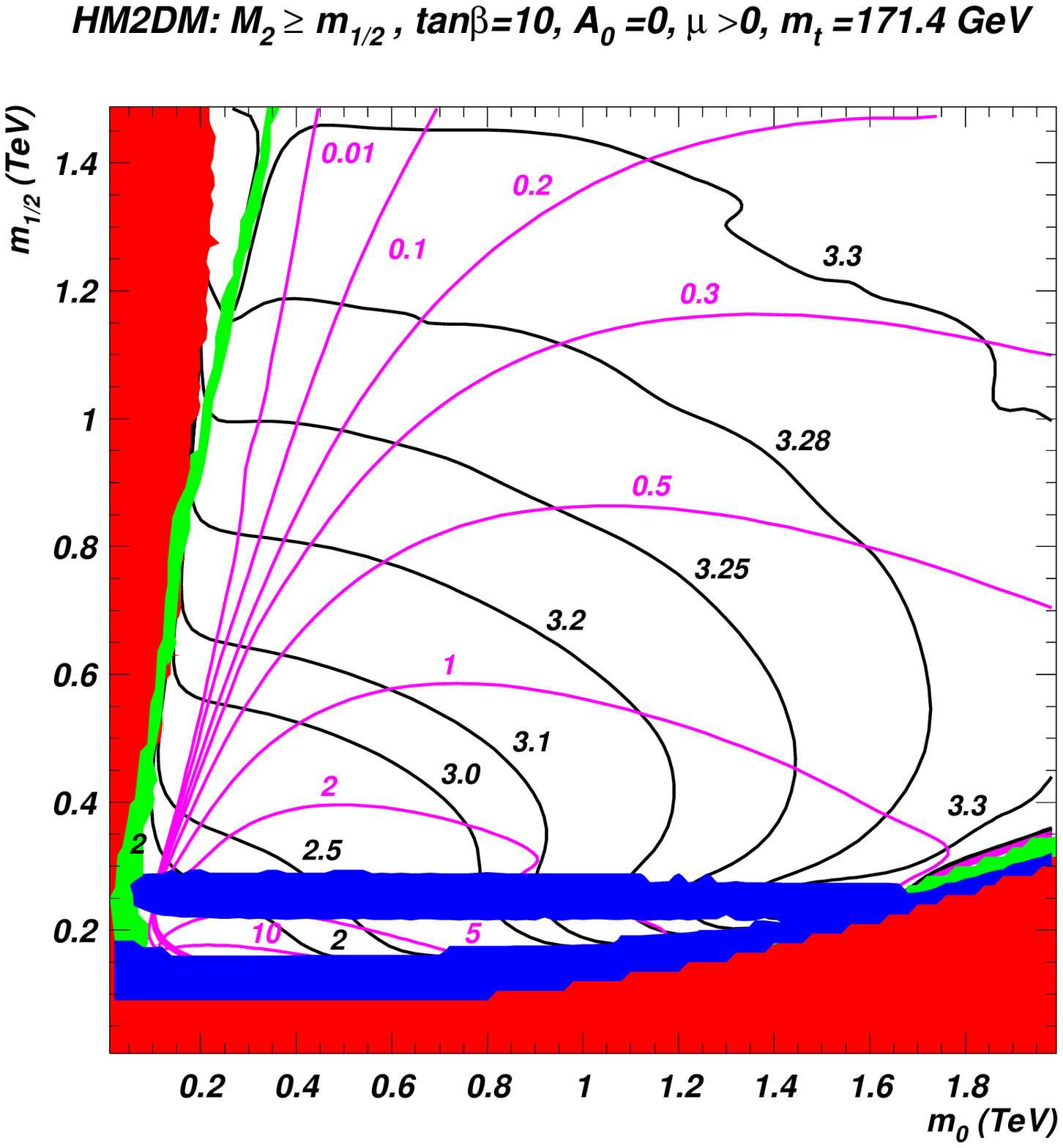,width=7cm,angle=0}
\epsfig{file=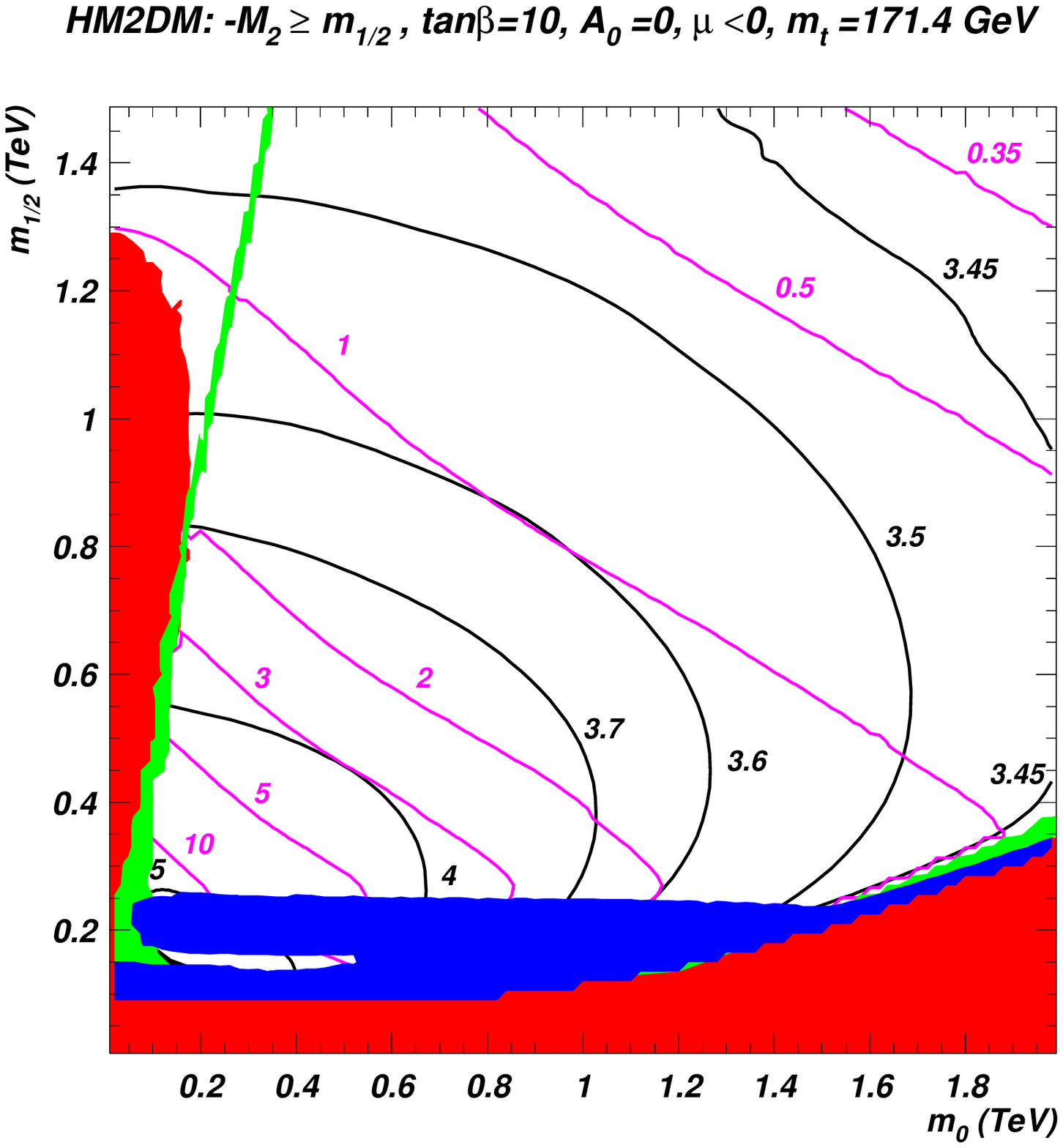,width=7cm,angle=0} }
\caption{\label{fig:bsg2} Contours of branching fraction $BF(b\to
s\gamma )\times 10^4$ (black) and $\Delta a_{\mu}\times 10^{10}$ (purple) 
in the $m_0\ vs.\ m_{1/2}$ plane for $A_0=0$, $\tan\beta
=10$, and {\it a})~$r_2>0$ with $\mu >0$, and 
{\it b})~$r_2<0$ with $\mu < 0$
 where $M_2$ has been dialed at every point to
large values such that $\Omega_{\tz_1}h^2\sim 0.1$.}}

\subsection{$(g-2)_\mu$}

Current measurements of the muon anomalous magnetic moment show an apparent
deviation from SM predictions. Combining QED, electroweak, hadronic
(using $e^+e^-\to {\rm hadrons}$ to evaluate hadronic loop contributions)
and light-by-light contributions, and comparing against measurements
from E821 at BNL, a {\it positive} deviation in 
$a_\mu\equiv \frac{(g-2)_\mu}{2}$ of
\be
\Delta a_\mu =a_\mu^{exp} -a_\mu^{SM} =22(10)\times 10^{-10} 
\ee
is reported in the Particle Data Book\cite{pdb}, {\it i.e.} a
$2.2\sigma$ effect.\footnote{More recent analyses\cite{davier} report a
larger discrepancy if only electron-positron data are used for the evaluation
of the hadronic vacuum polarization contribution; the significance of
the discrepancy is, however, reduced if tau decay data are used for this
purpose.} 

One-loop diagrams with $\tw_i-\tnu_\mu$ and $\tz_i-\tmu_{1,2}$ in the
loop would give supersymmetric contributions to $a_\mu$, perhaps
accounting for the (rather weak, yet persistent) discrepancy with the
SM.  For our canonical point with $m_0=m_{1/2}=300$~GeV, we have checked
that even though $|M_2|$ is large, the {\it total} chargino contribution
to $\Delta a_{\mu}$ dominates the {\it total} neutralino contribution
even in the HM2DM model exactly as in the mSUGRA case. This is the case for
both signs of $M_2$.  For $M_2 > 0$ the total neutralino contribution,
though much smaller than the corresponding chargino contribution, is
relatively larger in the HM2DM case as compared with mSUGRA.

%
%

The purple curves in Fig.~\ref{fig:bsg2} are contours of $\Delta
a_\mu^{\rm SUSY}\times 10^{10}$ in the $m_0\ vs.\ m_{1/2}$ plane.  For
$M_2 > 0$ shown in frame {\it a}), in the portion of the plane not
strongly excluded by the $b\to s\gamma$ constraint, we see that model,
$\Delta a_\mu^{\rm SUSY}$ is very small.  The situation for $M_2<0$ is
very similar as can be seen from frame {\it b}).  While it appears that
the HM2DM model will be strongly disfavoured if the muon magnetic moment
discrepancy continues to persist, we should remember that (1)~$\Delta
a_{\mu}^{\rm SUSY}$ and $BF(b\to s\gamma)$ are both sensitive to
$\tan\beta$, and (2)~the latter is very sensitive to small flavour violations
in the soft-SUSY breaking parameters which will not have any significant
effect on direct searches for supersymmetry.

%
%

\section{Direct and indirect detection of neutralino CDM}
\label{sec:dm}

In this section, we explore the prospects for direct and indirect
detection of neutralino dark matter within the HM2DM
framework\cite{eigen}.  
We adopt the IsaReS code\cite{isares} (a part of the Isatools package) for
the computation of the direct detection rates 
and the DarkSUSY code \cite{darksusy},
interfaced to Isajet, for the computation of the various indirect
detection rates. 
Indirect detection rates are sensitive to the DM distribution in our
galactic halo, larger rates being obtained for a clumpy or cuspy halo
distribution as compared with a smooth or less peaked distribution of the DM.  
We show our results for two halo profiles. The Adiabatically Contracted
N03 Halo model\cite{n03}, where the deepening of the gravitational
potential wells caused by baryon in-fall leads to a higher concentration
of DM in the center of the the Milky Way, gives higher detection rates,
especially for gamma ray and anti-particle detection than smoother halo
profiles.  For comparison, we also show projections using the Burkert
profile\cite{burkert} where the central cusp in the DM halo is smoothed
out by significant heating of cold particles.\footnote{For a comparison
of the implications of different halo model choices for indirect DM
detection rates, see {\it e.g.}  Refs.~\cite{bo,bbko,antimatter,nuhm}.}
In our analysis, we consider signals from the following processes.

\begin{enumerate}
\item Relic neutralinos in our galactic halo can scatter from nuclei in the
material of underground cryogenic or noble liquid 
detectors designed to detect the
resulting nuclear recoil, leading to direct detection of the neutralino~\cite{direct}.  Although there is no positive signal to date, the most
stringent upper limit on the scattering cross section comes from the
XENON-10 collaboration~\cite{xenon10}, which obtained an upper limit
$\sigma(\tz_1 p) \alt 8\times 10^{-8}$~pb for $m_{\tz_1}\sim 100$~GeV,
corresponding to the expected neutralino mass in the HM2DM model for our
canonical choice of parameters in Fig.~\ref{fig:hi_ino}.  We compute the
spin independent neutralino-proton scattering cross section (used as the
figure of merit in these experiments), and compare
it to projections for the sensitivity 
of Stage~2 detectors (CDMS2\cite{cdms2}, Edelweiss2\cite{edelweiss},
CRESST2\cite{cresst}, ZEPLIN2\cite{zeplin}) which are expected to probe
a factor of $\sim 5$ below the XENON-10 bound.\footnote{In our analysis,
we took the $\pi$-nucleon $\Sigma$ term to be 45~MeV. 
Different values recently suggested by other groups can change our
predictions by about factor of three~\cite{Ellis:2005mb}.}

 We also compare expectations in the HM2DM
model with the projected sensitivity of the proposed  
SuperCDMS detector with 25~kg of Ge, and with proposed ton-size noble liquid
detectors (XENON\cite{xenon}, LUX, WARP\cite{warp} and
CLEAN\cite{clean}), for which we use the sensitivity of Warm Argon
Project, with 1400~kg of argon as the benchmark.

\item Neutralinos gravitationally trapped in the core of the
sun may be indirectly detected
via their annihilation to
neutrinos at neutrino telescopes~\cite{neut_tel}.  Here, we present rates
for detection of $\nu_\mu \to \mu$ conversions at Antares\cite{antares}
or IceCube\cite{icecube}.  The reference experimental level we use is
the ultimate sensitivity of IceCube, with a muon energy threshold of 50~GeV, 
corresponding to a flux of about 40 muons per ${\rm km}^2$ per year.

\item Indirect detection of neutralinos may also be accomplished by
detection of high energy gamma rays from neutralino annihilation in
the galactic center. Such gamma rays\cite{gammas} have already been
detected by EGRET\cite{egret}, and will be searched for by the GLAST
experiment\cite{glast}.  We evaluate the integrated continuum $\gamma$
ray flux above a $E_\gamma=1$~GeV threshold, and take the GLAST
sensitivity of 1.0$\times10^{-10}\ {\rm cm}^{-2}{\rm s}^{-1}$ as our
benchmark.

\item Indirect detection of neutralinos is also possible via the
detection of anti-particles from neutralino annihilations in the
galactic halo. Proposed and on-going experiments include searches for
positrons\cite{positron} (HEAT\cite{heat}, Pamela\cite{pamela} and
AMS-02\cite{ams}), antiprotons\cite{pbar} (BESS\cite{bess}, Pamela,
AMS-02) and anti-deuterons ($\bar{D}$) (BESS\cite{bessdbar}, AMS-02,
GAPS\cite{gaps}).  For positrons and antiprotons we evaluate the
averaged differential antiparticle flux in a projected energy bin
centered at a kinetic energy of 20~GeV, where we expect an optimal
statistics and signal-to-background ratio at space-borne antiparticle
detectors\cite{antimatter,statistical}. 
We take the experimental
sensitivity to be that of the Pamela experiment after three years of
data-taking as our benchmark.  Finally, we evaluate the average
differential anti-deuteron flux in the $0.1<T_{\bar D}<0.25$~GeV range,
where $T_{\bar D}$ stands for the anti-deuteron kinetic energy per
nucleon, and compare it to the estimated GAPS sensitivity for an
ultra-long duration balloon-borne experiment~\cite{gaps} (see
Ref.~\cite{baerprofumo} for an updated discussion of the role of
antideuteron searches in DM indirect detection).

\end{enumerate}

In Fig.~\ref{dmrates}, we illustrate the various direct and indirect DM
detection rates for our canonical case with $m_0=m_{1/2}=300$~GeV,
$A_0=0$, $\tan\beta =10$ and $\mu >0$, where $M_2$ is allowed to vary.
The $M_2$ value corresponding to the mSUGRA model is denoted by a 
solid black vertical line at $r_2=1$,
 while the HM2DM scenarios for $r_2<0$ and $r_2>0$ with
$\Omega_{\tz_1}h^2$ within the WMAP range (\ref{wmap})
are shown by the green regions on the left and
right ends of the plot. The MWDM and BWCA solutions are seen as the
very narrow green regions near $r_2 \sim \pm (0.5-0.6)$.
The dotted lines correspond to the sensitivity level for representative
searches: {\it i.e.}, the signal is observable only when the model
prediction is higher than the corresponding dotted line.

In frame~{\it a}), we show the spin-independent neutralino-proton
scattering cross section.  We see that for a bino-like
neutralino with $m_{\tz_1}\sim 100$~GeV as in the  mSUGRA or BWCA cases,
the signal will only be
detectable if the cross section can be probed at the 10$^{-9}$~pb level
as envisioned at superCDMS or at 100-1000~kg noble liquid detectors.
In contrast, in both the HM2DM regions where the neutralino has a significant
higgsino component, the cross section is 
just below the current bound, and should be detectable at
CDMS2. 
This is simply a reflection of the
well-known result that MHDM has rather large neutralino-proton
scattering rates, as is typified by the HB/FP region of the mSUGRA
model, and further,  that experimental sensitivity at the $10^{-8}$~pb level
will probe a wide class of models with MHDM~\cite{wtn_dm}.

\FIGURE[tbh]{
\mbox{\hspace{-1cm}
\epsfig{file=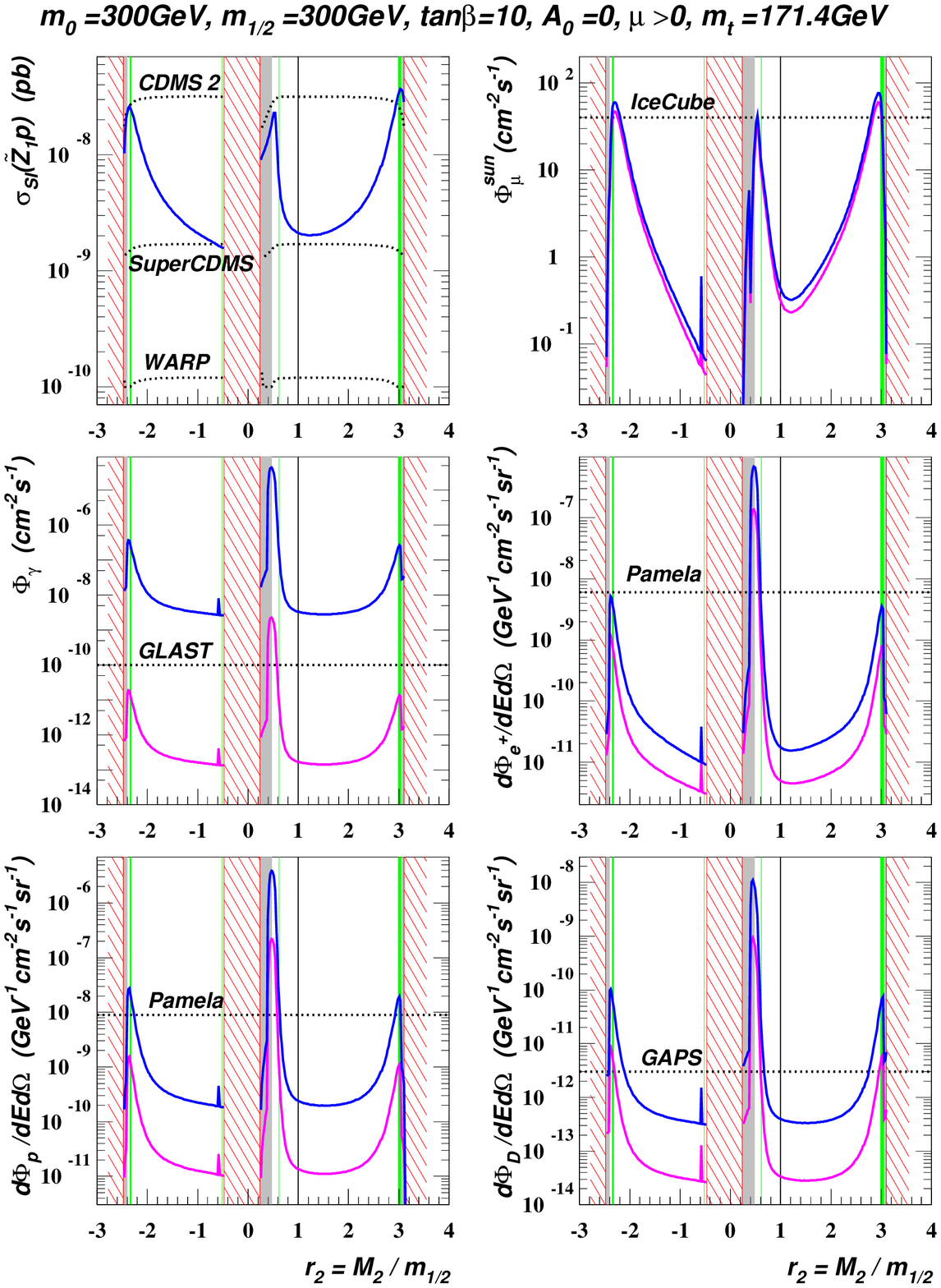,width=12cm,angle=0} }
\caption{\label{dmrates}
Rates for direct and indirect detection of neutralino dark matter
{\it vs.} $r_2$ for $m_0=m_{1/2}=300$~GeV, with
$\tan\beta =10$, $A_0=0$, $\mu >0$. 
The various shadings are the same as in Fig.~\ref{fig:rd}. 
The blue curves correspond to the Adiabatically Contracted N03 dark matter halo
model, while the purple ones are for the Burkert profile. For each
experiment, the signal is
observable if the rate is above the corresponding
dotted curve.}}

In Fig.~\ref{figdotplot}, we show the expected value of
$\sigma_{SI}(\tz_1 p)$ in the HM2DM model, resulting from a scan in
$m_0$ and $m_{1/2}$, keeping $\tan\beta=10$ and $A_0=0$, and where we
adjust $M_2$ to get agreement with (\ref{wmap}), for $M_2>0$ with
$\mu>0$ (upper frame) and $M_2< 0$ with $\mu < 0$ (lower frame). Also
shown are the sensitivity limits for CDMS2, superCDMS and WARP
1400~kg. The most striking feature of the figure is that the bulk of the
points in the scan give a cross section around $10^{-8}$~pb, independent
of the neutralino mass. This is because the increased bino-higgsino
mixing necessary to maintain agreement with (\ref{wmap}) for larger
values of $m_{\tz_1}$ compensates for the drop in cross section for
larger neutralino masses.  We also see that the cross sections for
negative $M_2$ are systematically lower than those for $M_2>0$. We have
checked that flipping the relative sign between $M_1$ and $\mu$ causes a
flip in the relative sign between the $h\tz_1\tz_1$ coupling and the
$H\tz_1\tz_1$ coupling, so that $h$- and $H$-mediated amplitudes for
neutralino-nucleon scattering interfere constructively in the positive
$\mu$ case and destructively for negative $\mu$, accounting for the drop
in the cross section. (Since squarks are heavy, squark-mediated
amplitudes are negligible.)\footnote{In more detail, the flip of the
sign of $\mu$ flips the relative sign between the up- and down- higgsino
components in $\tz_1$.  Further, since $m_A \gg M_{\rm weak}$ for the
HM2DM model, the Higgs mixing angle satisfies $\tan\alpha \simeq
\cot\beta$, so that for $\tan\beta\agt 10$, $Re{h_u^0}\sim h$ and
$Re{h_d^0}\sim H$. As a result, there is a flip of the relative sign
between the $\tz_1\tz_1 h$ and $\tz_1\tz_1 H$ couplings.}  The cluster
of points below $m_{\tz_1}=100$~GeV in the upper frame are points in the
white strip at $m_{1/2}\sim 0.2$~TeV inside the blue region in
Fig.~\ref{fig:r2}{\it a}). We see that while superCDMS should probe all
the points for positive values of $M_2$, the increased sensitivity of
ton-size noble element detectors appears essential in the $M_2< 0$ case.

\FIGURE[tbh]{
\mbox{\hspace{-1cm}
\epsfig{file=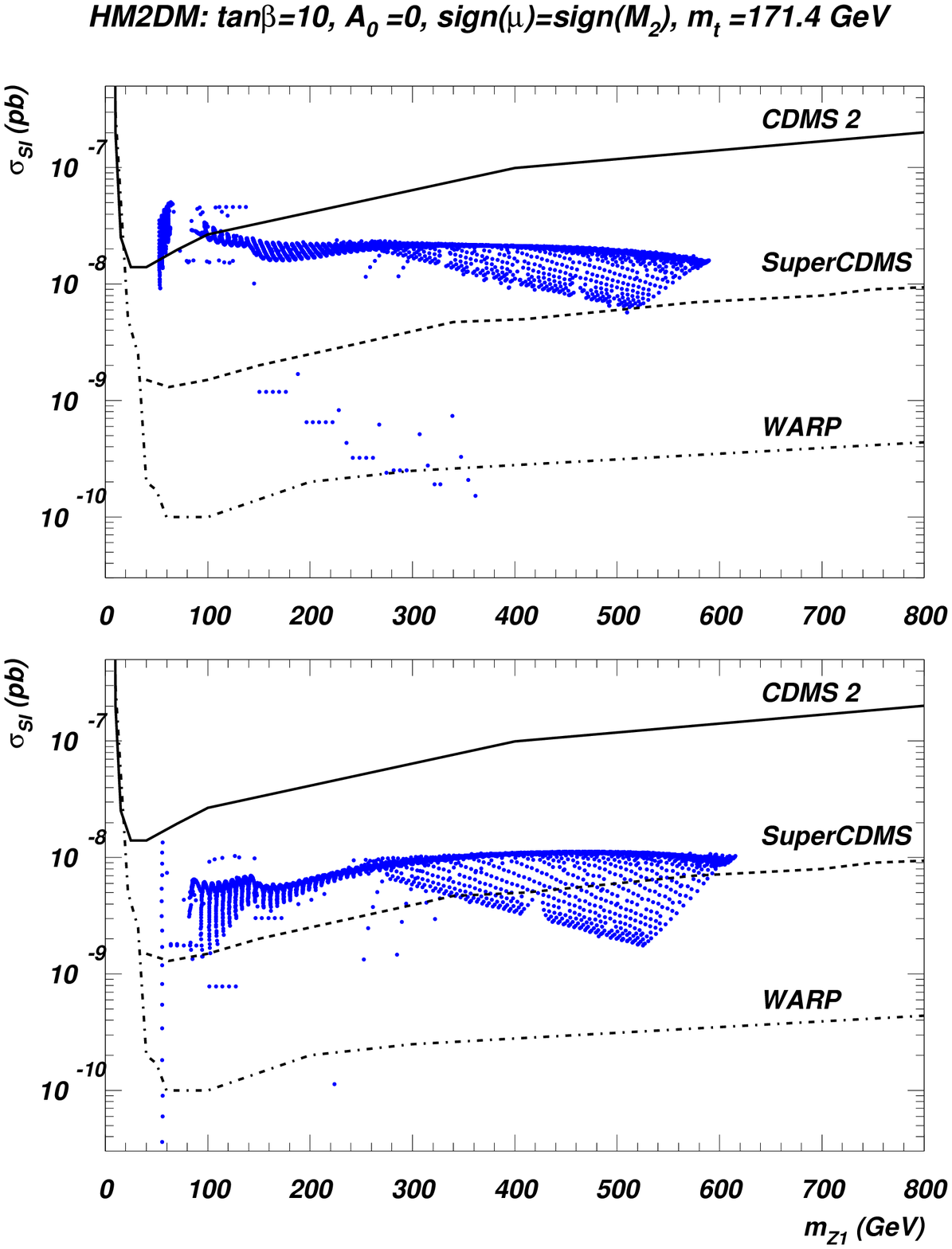,width=12cm,angle=0} }
\caption{\label{figdotplot} The neutralino-proton scattering cross
  section for the HM2DM model with  $\tan\beta=10$ and $A_0=0$, but
  where we scan over $m_0$ and $m_{1/2}$ for $M_2 >0$ with $\mu>0$
  (upper frame),  and $M_2 <0$ with $\mu<0$ (lower frame). For each
  point on this plot, $M_2$ is adjusted to saturate the observed DM
  density. We retain only those points where $m_{\tw_1} \ge 103.5$~GeV
  in this plot.
  }}

Turning to indirect dark matter detection, in Fig.~\ref{dmrates}{\it
b}), we show the flux of muons from neutralino pair annihilations in the
core of the Sun, again for our canonical choice of parameters introduced
in Fig.~\ref{fig:hi_ino}. The blue and purple curves are our projections  
assuming the Adiabatically Contracted N03 Halo
model\cite{n03} and the Burkert profile\cite{burkert}, respectively, 
for the distribution of DM in our galaxy.
In this case, the result is 
only mildly sensitive to the choice of halo distribution, since the muon
flux is mainly determined by the equilibrium density of neutralinos in
the sun, and the cross section for neutralino annihilation into
neutrinos. We see that the expected muon flux is more than an order of
magnitude below the projected sensitivity of IceCube in the mSUGRA
framework, but is larger by over two orders of magnitude, and in the
observable range, for the HM2DM model on account of the increased
higgsino content of the LSP.  The muon flux rapidly drops off near the
ends where $m_{\tz_1}$ falls below $M_W$ because the neutrinos (which
then mainly come from decays of heavy quarks) become very soft, so that their
efficiency for detection at IceCube is vastly degraded. 


In frames {\it c}), {\it d}), {\it e}) and {\it f}) we show the flux of
photons, positrons, antiprotons and antideuterons, respectively.  Also
shown by the horizontal lines are the anticipated experimental
sensitivities. Again, we show results using the Adiabatically Contracted
N03 Halo model (blue) and the Burkert profile (purple). We see that the
projections are sensitive to the assumed halo distribution. This is most
striking for the photon signal at GLAST, where the difference is more
than four orders of magnitude. This is because most of the photons come
from the galactic center where the difference between the two profiles
is the most pronounced. In contrast, projections for the detection of
positrons, anti-protons and anti-deuterons from neutralino annihilation
(unlike photons, these can reach the earth only from limited distances)
differ by a factor of 5-15.  Again, the rates for indirect detection via
observation of halo annihilation remnants, which are typically low for
bino-like DM as in the mSUGRA model, jump by a factor of 30-300 in the
HM2DM model where $|r_2|$ is increased so that the measured CDM relic
abundance is obtained. For our choice of model parameters, the GAPS
experiment should be able to detect anti-deuterons even for the case of
the Burkert halo profile, while the positron signal in Pamela is
projected to be just below its sensitivity limit even for the optimistic N03
Halo profile. The situation for the anti-proton signal is less
conclusive since its detectability clearly depends on the halo
distribution. 

\section{Supersymmetry signals at colliders}
\label{sec:col}

We now turn to an examination of the implications of the
HM2DM model for SUSY collider searches at the Fermilab Tevatron, 
the CERN LHC, and a 0.5-1.5~TeV linear $e^+e^-$ collider.
The sparticle mass spectrum in the HM2DM model qualitatively
differs from mSUGRA
in several respects: {\it i}) the low $|\mu |$ parameter
implies that charginos and neutralinos should be lighter
than in mSUGRA cases with gaugino mass unification and a large
$|\mu |$ parameter, so these sparticles should be more accessible to collider
searches. In addition, {\it ii}) 
the large $|M_2|$ parameter means $\tw_2$ and $\tz_4$
will be quite heavy and nearly pure wino states, so likely 
difficult to access at colliders, except perhaps via the (kinematically
suppressed) production and
subsequent decays of $\tq_L$. Finally, {\it iii}) the large $|M_2|$ 
parameter pushes left-sfermion soft terms to higher values, so that 
the lighter sfermions are dominantly right-sfermions.

In Fig.~\ref{fig:wino}, we show contours of $m_{\tw_1}$ in the $m_0\
vs.\ m_{1/2}$ plane for $A_0=0$, $\tan\beta =10$ and $\mu >0$, where at
every point $M_2$ has been dialed up to obtain MHDM with
$\Omega_{\tz_1}h^2\sim 0.1$. We see that throughout the plane,
$m_{\tw_1}\sim {1\over 2} m_{1/2}$, whereas in mSUGRA, $m_{\tw_1}\sim
{2\over 3} m_{1/2}$.  This may be of relevance at an $e^+e^-$ collider
where the determination of both $m_{\tw_1}$ and $m_{1/2}$ (via the
determination of $m_{\tz_1}$) along with $M_2$ may be possible if
$\tw_1\tw_2$ production is kinematically accessible.

\FIGURE[tbh]{
\mbox{\hspace{-1cm}
\epsfig{file=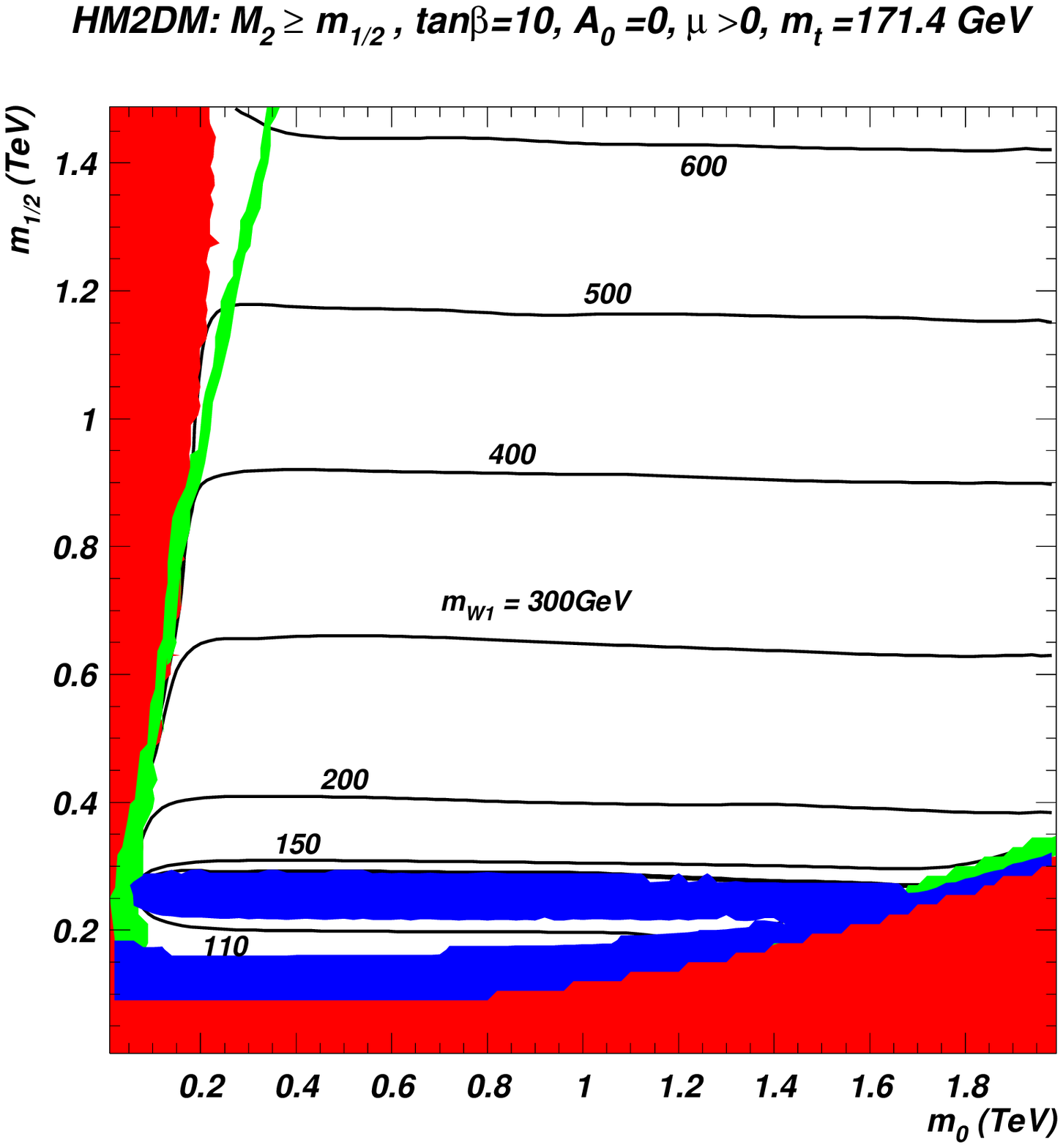,width=12cm,angle=0}}
\caption{\label{fig:wino}
Contours of $m_{\tw_1}$ in the $m_0\ vs.\ m_{1/2}$ plane for
$\tan\beta =10$, $A_0=0$ and $\mu >0$ where $M_2$ has been dialed
up at every point to yield MHDM where $\Omega_{\tz_1}h^2\sim 0.1$. 
}}

In addition, in the HM2DM model, since $\tz_2$ and $\tz_3$ contain large
higgsino components while $\tz_1$ is a mixed higgsino-bino state, the
mass gaps $m_{\tz_3}-m_{\tz_1}$ and $m_{\tz_2}-m_{\tz_1}$ will be
expected to be much smaller than in mSUGRA. In mSUGRA,
$m_{\tz_2}-m_{\tz_1}\sim 0.4 m_{1/2}$ so that as $m_{1/2}$ grows, the
growing $\tz_2 -\tz_1$ mass gap ultimately allows the $\tz_2\to\tz_1 Z$ or
$\tz_1 h$ two-body decays to turn on, which dominate the $\tz_2$ branching
fraction. The two-body ``spoiler'' decay modes\cite{trilep} turn off the
leptonic decays $\tz_2\to\tz_1\ell\bar{\ell}$, which can be the starting
point for sparticle mass reconstruction in gluino and squark cascade
decays\cite{cascade} at hadron colliders.  In Fig.~\ref{fig:mz2mz1}{\it
a}), we show the $\tz_2 -\tz_1$ mass gap in the HM2DM model in the same
plane as in Fig.~\ref{fig:wino}.  We see that in the case of HM2DM
model, the mass gap is everywhere less than $M_Z$, so that
$\tz_2\to\tz_1 \ell\bar{\ell}$ decays will not be shut off by the two-body 
spoiler modes. Moreover, the mass gap is also almost always larger
than $\sim 25$~GeV, and decreases with increasing $m_{1/2}$ in contrast
to the cases of MWDM and DM via BWCA, where the mass gap increases to
beyond 100~GeV for the largest values of $m_{1/2}$. This could serve to
distinguish HM2DM from other scenarios, something we will return to
below.
Also, in contrast to mSUGRA as well as the MWDM and BWCA frameworks, in the
HM2DM model (and other models with MHDM), we expect that
$m_{\tz_3}-m_{\tz_1}$ cannot be too large because the mass of
$\tz_3$ is expected to be 
not very far above $|\mu|$, as we saw in Fig.~\ref{fig:mass}. In
Fig.~\ref{fig:mz2mz1}{\it b}), we show contours of
$m_{\tz_3}-m_{\tz_1}$. We see that this difference is also everywhere 
smaller than $M_Z$, so that the decays $\tz_3\to \tz_1\ell\bar{\ell}$ and 
$\tz_3\to \tz_2\ell\bar{\ell}$ are not shut off by the spoiler
modes. 

\FIGURE[tbh]{
\mbox{
\epsfig{file=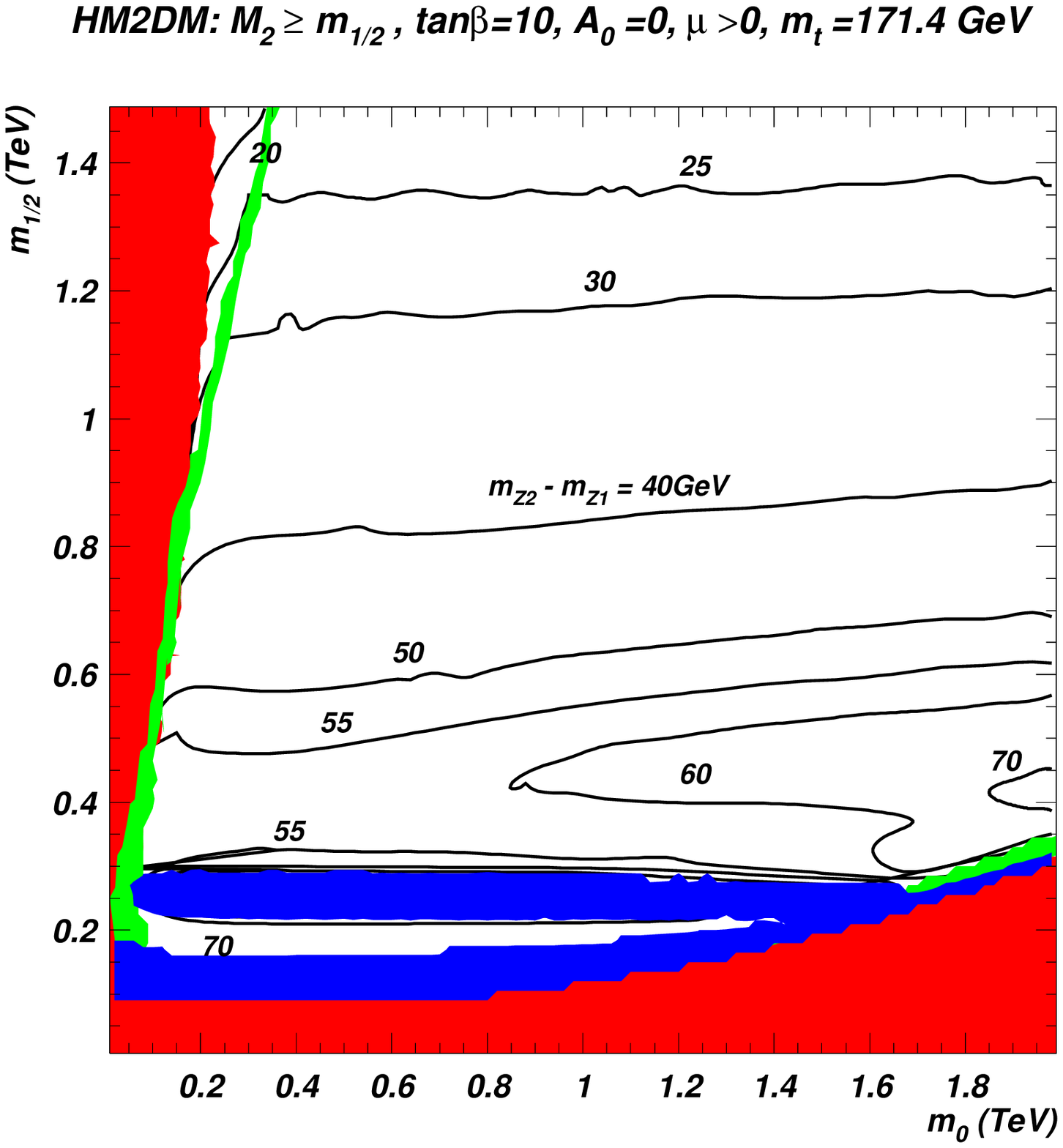,width=7cm,angle=0} 
\epsfig{file=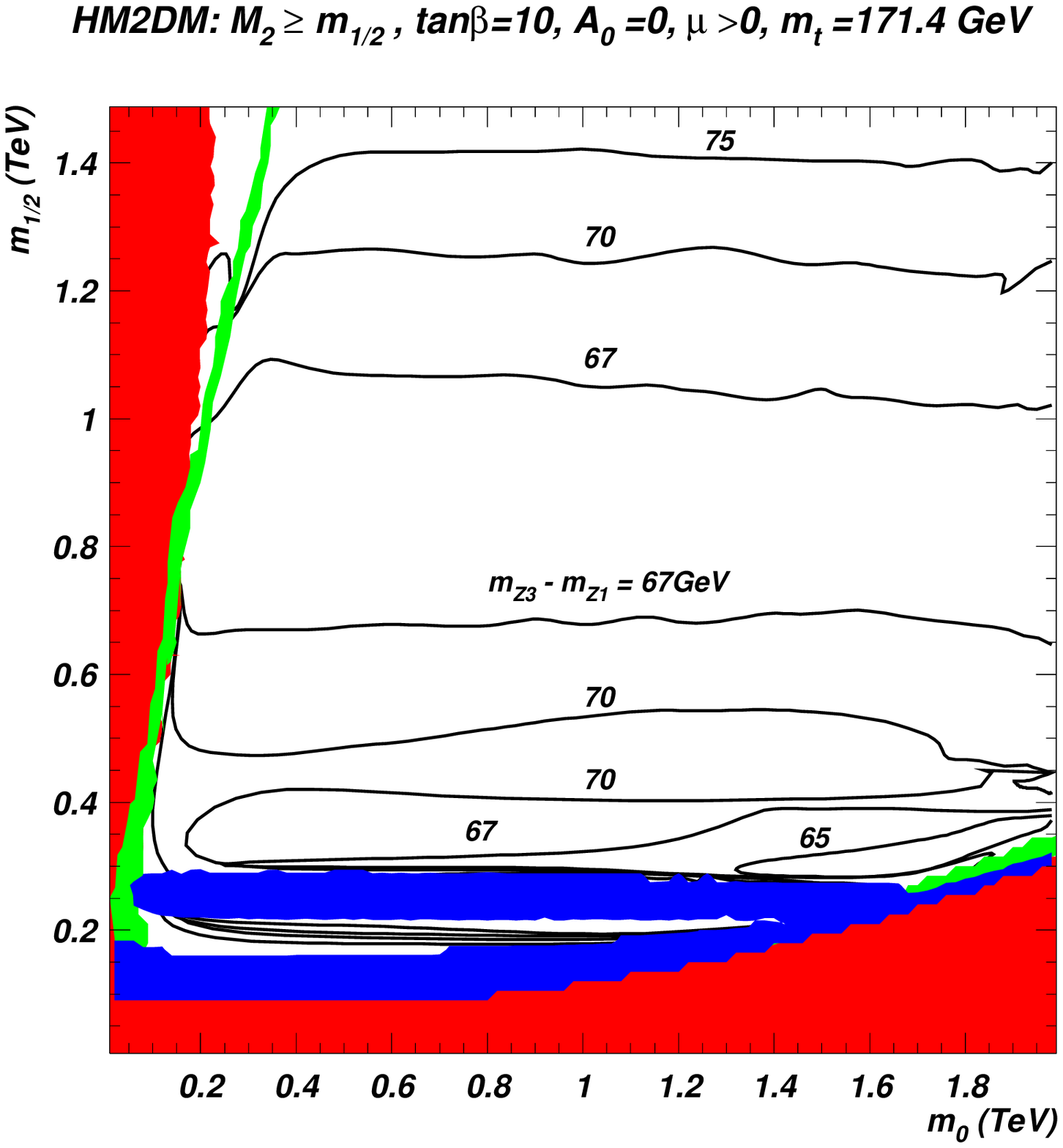,width=7cm,angle=0} }
\caption{\label{fig:mz2mz1}
Contours of {\it a})~$m_{\tz_2}-m_{\tz_1}$, and {\it
  b})~$m_{\tz_3}-m_{\tz_1}$, in the $m_0\ vs.\ m_{1/2}$ plane for
$\tan\beta =10$, $A_0=0$ and $\mu >0$ where $M_2$ has been dialed
up at every point to yield MHDM where $\Omega_{\tz_1}h^2\sim 0.1$. 
}}

\subsection{Fermilab Tevatron}
\label{ssec:tev}

In the mSUGRA model, since $m_{\tw_1}>103.5$~GeV from LEP2 searches, we
expect 
 $m_{\tg} \simeq 3.5 m_{\tw_1}\agt 350-400$~GeV,
and this high of
sparticle masses generally gives quite low $\tg\tg$, $\tq\tq$ and
$\tg\tq$ production cross sections\cite{tevreach}.  Gluino and
right-squark masses are relatively unchanged in going from mSUGRA to
HM2DM, while left-squark masses typically increase.  Thus we expect that
gluino and squark production rates at the Tevatron will be relatively
low, and the signal not easily extracted from data in the HM2DM model.

Another possibility is to look for $p\bar{p}\to \tw_1\tz_2\to
3\ell+\eslt$\cite{trilep} in the HM2DM model. The large higgsino components of
$\tw_1$ and $\tz_2$ imply that $\tw_1\tz_2$ production via the $W^*$
will dominantly occur via the isodoublet couplings of the neutralinos to
the $W$. However, this (iso-doublet) $W\tw_1\tz_2$ coupling will now be
smaller than the corresponding coupling in the mSUGRA framework (where
the coupling arises from the iso-triplet components of $\tw_1$ and
$\tz_2$, and so is very large) so that $\sigma(\tw_1\tz_2)$ 
will be suppressed in HM2DM relative to mSUGRA.

We are thus led to
re-examine the rate for trilepton production at the Tevatron in HM2DM
model and compare this to the mSUGRA case where the trilepton plus
$\eslt$ signal is regarded as the gold-plated signature.  We use the
cuts SC2 proposed in Ref.~\cite{new3l} which allow for efficient
extraction of the $3\ell+\eslt$ signal from various SM backgrounds, the
largest of which are $W^*Z^*\to 3\ell+\nu$ and $W^*\gamma^*\to 3\ell
+\nu$ production.  In Fig.~\ref{fig:sig3l}, we plot the $3\ell$ cross
section after cuts SC2, fixing $m_0=300$~GeV, $A_0=0$, $\tan\beta =10$
and $\mu >0$ versus $m_{\tw_1}$, obtained by varying $m_{1/2}$.
Contrary to expectation, we see that the signal is larger in the HM2DM
framework than in the mSUGRA case. We have traced this to the fact that
the leptonic decay of $\tz_2$ is very suppressed within the mSUGRA
framework, while the corresponding branching fraction is just under
$B(Z^0 \to \ell\bar{\ell})\simeq 6\%$ as expected for the case of MHDM.
In the mSUGRA case, the signal rate is always below the 0.8~fb
level. The total background estimated in Ref. \cite{new3l} is 1.05~fb,
which requires a 1.6~fb signal to give a $5\sigma$ effect with 10~fb$^{-1}$ of integrated luminosity.  The HM2DM curve is enhanced
relative to mSUGRA, and reaches a maximum of about 1.4~fb -- just below
the edge of observability.  The dashed region between
$m_{\tw_1}\sim 117-137$~GeV is excluded as can be seen from Fig.~\ref{fig:r2},
while for higher $m_{\tw_1}$ values, the signal is always below 0.6~fb,
so will be difficult to extract at the Tevatron.
\bigskip
\bigskip
\bigskip

\FIGURE[tbh]{
\mbox{\hspace{1cm}
\epsfig{file=sigma3l.eps,width=12cm,angle=0} }
\caption{\label{fig:sig3l} Trilepton signal after cuts SC2 from
Ref. \cite{new3l} versus $m_{\tw_1}$ along a line of $m_0=300$~GeV,
variable $m_{1/2}$, adjusted to reproduce the value of $m_{\tw_1}$,
$\tan\beta =10$, $A_0=0$ and $\mu >0$ where $M_2$ has been dialed up at
every point to yield MHDM where $\Omega_{\tz_1}h^2\sim 0.1$.  We also
show the corresponding rate from mSUGRA model where $M_2=m_{1/2}$, and
the $5\sigma$ discovery level for 10~fb$^{-1}$ of integrated
luminosity. The dashed region with 117~GeV $\alt m_{\tw_1} \alt$ 137~GeV
is excluded in the HM2DM model as can be seen from Fig.~\ref{fig:r2}.
}}

\subsection{CERN LHC}
\label{ssec:lhc}

At the CERN LHC, gluino and squark pair production is the dominant
sparticle production mechanism if sparticle masses are in the TeV 
range~\cite{wss}. Gluino and squark cascade decays give rise to 
events containing multiple hard jets, isolated leptons, $\eslt$, and
sometimes also isolated photons.
The reach of the LHC has been calculated in the mSUGRA model in 
Refs.~\cite{lhcreach}. The ultimate reach in the $\eslt + {\rm jets}$ 
channel is relatively insensitive to the details of the cascade decays, 
but depends mostly on the gluino and squark production 
cross sections, which in turn depends only on their masses. 

We have made a rough translation of the LHC reach in the mSUGRA framework to
that of the HM2DM model by computing the total production cross section
along the 100~fb$^{-1}$ reach contour in the last paper of
Ref.~\cite{lhcreach}, and equating it to total production cross sections
in the HM2DM $m_0\ vs.\ m_{1/2}$ plane. The result is shown as the
dotted contour labelled LHC in Fig.~\ref{fig:lhcilc}. At the small $m_0$
end, the contour ends up being typically lower in $m_{1/2}$ values by
about 150~GeV than the corresponding plots in mSUGRA, with the reduction
being smaller for larger values of $m_0$. This is because in the HM2DM
case, for given $m_0$ and $m_{1/2}$ values, the various left-squark
masses are raised up by a few hundred GeV, causing the corresponding
production cross sections to drop. Thus, somewhat lower gluino and
right-squark masses are needed to obtain the same reach level as mSUGRA
within the HM2DM framework. The location
of the LHC reach contour in the HM2DM model varies between
$m_{1/2}\sim 1260$~GeV at low $m_0$ and $m_{1/2}\sim 1040$~GeV at $m_0=2$~TeV.  This corresponds to a reach in terms of gluino mass of
$m_{\tg}\sim 2350-2750$~GeV for 100~fb$^{-1}$ of integrated luminosity.
\FIGURE[tbh]{
\mbox{\hspace{-1cm}
\epsfig{file=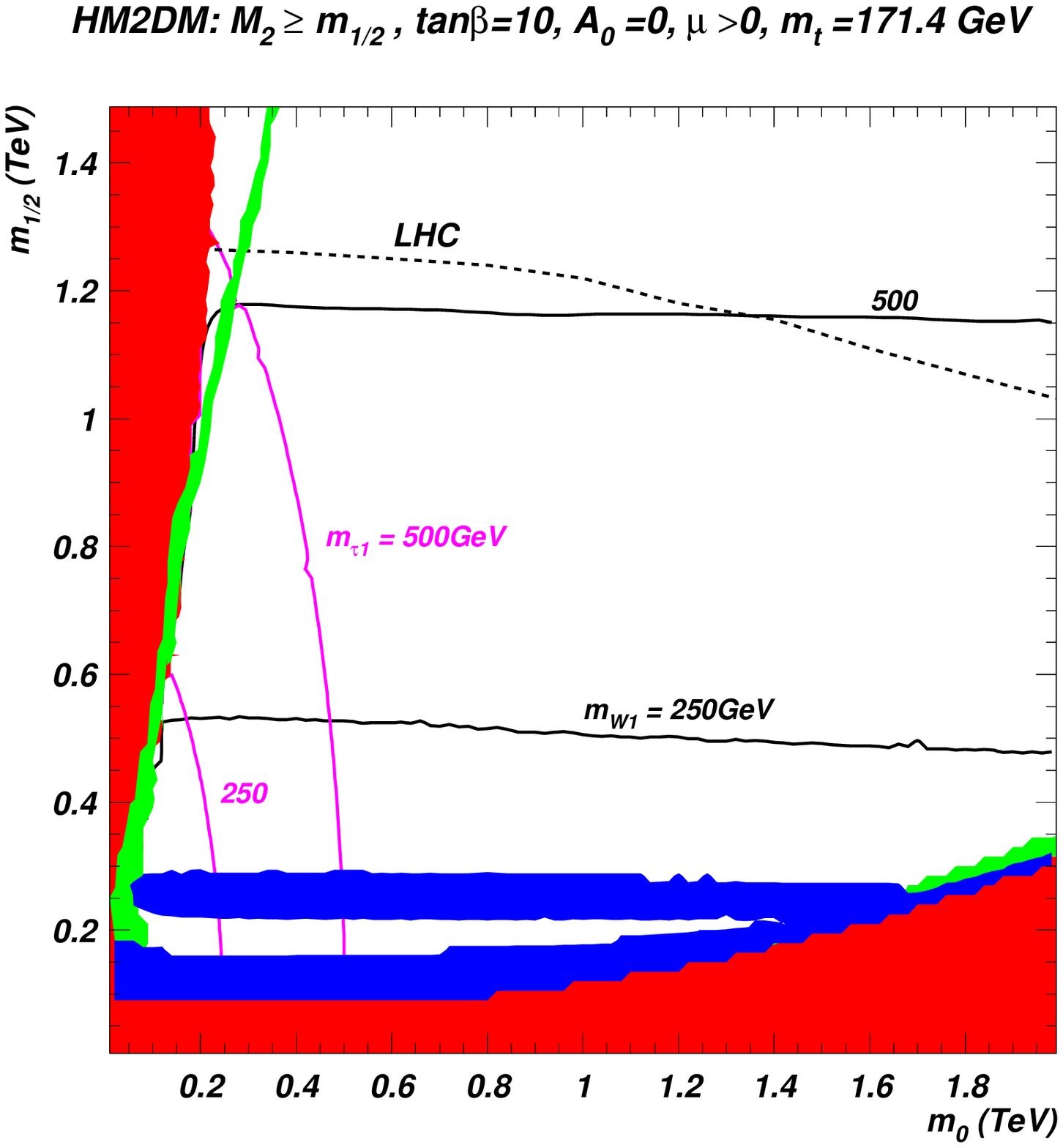,width=12cm,angle=0} }
\caption{\label{fig:lhcilc}
Approximate reach of the CERN LHC (with 100~fb$^{-1}$ of data)
and ILC with $\sqrt{s}=0.5$ and 1~TeV in the HM2DM model,
viewed in the $m_0\ vs.\ m_{1/2}$ plane for
$\tan\beta =10$, $A_0=0$ and $\mu >0$ where $M_2$ has been dialed
up at every point to yield MHDM where $\Omega_{\tz_1}h^2\sim 0.1$. 
}}

If supersymmetry is discovered at the LHC, reconstruction of SUSY events
to measure sparticle masses will be an important item on the experimental
agenda. Edges in the distribution of opposite sign, same flavour
dilepton masses provide important information on the mass difference
$m_{\tz_i}-m_{\tz_j}$ between those neutralino pairs for which the three
body decay $\tz_i \to \tz_j\ell\bar{\ell}$ has a significant branching
fraction~\cite{mlledge}. In the mSUGRA framework, we typically expect an
observable edge just for the decays of
$\tz_2$, if its mass is low enough so that the spoiler two body modes are
not accessible. However, for the HM2DM model, as for all models with
MHDM, both $\tz_2$ and $\tz_3$ are relatively light, so that we may
expect more than one mass edge in the dilepton mass spectrum.

To examine this, we simulated 1M LHC SUSY events for the HM2DM1 point
in Table~\ref{tab:m2dm} using Isajet~7.76, and passed these through a toy
detector simulation as described in Ref.~\cite{gabe}. In addition to the
various geometric and other acceptance requirements, we required the
following analysis cuts on the signal:
$$\eslt > (100~{\rm GeV}, 0.2M_{\rm eff}),$$
$$n_{\rm jets}\ge 4,$$
$$E_T(j_1, j_2, j_3, j_4) \ge (150, 150, 80, 50)~{\rm GeV},$$
$$S_T\ge 0.2. $$ For the present analysis, we only retain events which
include, in addition, a pair of opposite sign (OS) dileptons.  The
invariant mass distribution for $e^+e^- +\mu^+\mu^-$ pairs in SUSY
events is shown by the red-shaded histogram for the signal, and by the
line histogram for the SM background, in Fig.~\ref{fig:dilep}{\it a}). In
the signal, these leptons arise from the decays of neutralinos and
charginos produced in the SUSY cascade. The mass of dileptons from the
three body decay $\tz_i \to \tz_j \ell\bar{\ell}$ is kinematically
bounded by $m_{\tz_i}-m_{\tz_j}$, resulting in the mass edges mentioned
above. Lepton pairs where the leptons each come from decays of {\it
different} charginos or neutralinos are uncorrelated in mass, and so are
expected to yield a smooth broad continuum. Since these pairs are also
uncorrelated in flavour, subtracting the distribution of {\it
unlike-flavour} OS dilepton pairs should (up to statistical
fluctuations) leave us with the dilepton mass distribution of dileptons
from the same neutralino~\cite{sub}. This subtracted distribution is
shown in Fig.~\ref{fig:dilep}{\it b}). The following features are worth
noting.
\begin{enumerate}
\item The red shaded histogram in frame {\it a}) shows distinct gaps
  near around
  46~GeV and at around 64~GeV, close to the mass edges at 47.6~GeV
  and 65.4~GeV expected from the leptonic decays of $\tz_2$ and $\tz_3$
  to the LSP, respectively.   
 There is no corresponding gap near $m_{\tz_3}-m_{\tz_2}$
  indicating that the branching fraction $B(\tz_3\to
  \tz_2\ell\bar{\ell})$ is very small (0.2\% per lepton flavour in the
  present case).
\item There is a distinct peak at $M_Z$ showing that $Z$ bosons are
  being produced in SUSY cascades. This is a strong indication (even if
  not evidence) of the production of $\tw_2$ and/or $\tz_4$, since the only
  other source of $Z$'s would be the decays $\tst_2\to \tst_1 Z$,
  $\tb_2\to \tb_1Z$ or $A \to hZ$, all of which are likely to have very
  small cross sections in many models.
\item The subtracted distribution in frame {\it b}) also shows these edges
 even though the location of the $m_{\tz_2}-m_{\tz_1}$ edge seems
 somewhat smeared out by an upward fluctuation in the opposite flavour OS
 dilepton distribution in the mass bin just below 60~GeV, resulting
 instead in a shoulder around the expected value. 
\item The large levels of the background in some bins are caused by a
 handful of QCD events passing our cuts. Because of the large QCD cross
 section, we need to simulate a much larger sample of QCD events than we
 were able to in order to further reduce these fluctuations.\footnote{It is not clear that such a computer-intensive simulation of
 the tail of this detector-dependent background would be particularly
 meaningful, especially in view of
 our toy calorimeter simulation.}  We expect though that the QCD background
 to the SUSY  signal will be under control at te LHC.
\item Finally, we observe that in both frames the shape of the dilepton
  distribution in the region below the first edge, and in between the
  two edges is very different. Specifically, the $m_{\tz_3}-m_{\tz_1}$
  mass edge is much sharper than the $m_{\tz_2}-m_{\tz_1}$ edge. This is
  due to the difference in shapes of the mass distributions of dileptons
  from the decays of $\tz_2$ and $\tz_3$. 
  As pointed out by Kitano and Nomura in Ref. \cite{mllshape}, 
  this shape depends on the relative sign of the
  mass eigenvalue of the parent and daughter neutralino. If this
  relative sign is positive, we get a sharp mass edge as is the case for
  the $m_{\tz_3}-m_{\tz_1}$ edge in the figure. Since $\tz_2$ and
  $\tz_3$ are dominantly the higgsino states, not surprisingly they then
  have opposite signs for the eigenvalues, causing the distribution of
  dileptons from $\tz_2 \to \ell\bar{\ell}\tz_1$ to peak at lower values
  of $m(\ell^+\ell^-)$ so that the $m_{\tz_2}-m_{\tz_1}$ edge is much more
  diffuse. The shape of the dilepton distribution thus provides
  important information about the composition of neutralinos.
\end{enumerate}

\FIGURE[htb]{
\epsfig{file=hm2dm1-1M-SFOS_cutsC1a.eps,width=7.cm} \hspace{0.6cm}
\epsfig{file=hm2dm1-1M-OSdiff_cutsC1a.eps,width=7.cm}
\caption{\label{fig:dilep} {\it a})~Distribution of the invariant mass
of same flavor/opposite sign dilepton pairs in SUSY events at the CERN
LHC (red histogram) after the cuts discussed in the text, along with the
expectation for the corresponding SM background (open histogram). {\it
b}) The difference between this distribution and the corresponding one
with opposite flavour/opposite sign deleptons in SUSY events (blue
histogram) along with the SM background (open histogram). 
The signal distributions are calculated for the 
HM2DM1 case in Table \ref{tab:m2dm}.
\label{fig:mll}}}

\subsection{Linear $e^+e^-$ collider}
\label{ssec:ilc}

In assessing the role of the ILC for a discovery of SUSY within the
HM2DM framework, we first note that, for a fixed value of $m_0$ and
$m_{1/2}$, the value of $m_{\tw_1}$ is lowered with respect to mSUGRA.
This is because in the HM2DM model, $|\mu|$ is lowered to below $\sim 2M_1$, 
which is approximately the $\tw_1$ mass in models with unified gaugino masses.  This increases the reach of ILC
in $m_0\ vs.\ m_{1/2}$ space while at the same time the LHC reach is
slightly diminished due to the increased left-squark masses. We show in
Fig. \ref{fig:lhcilc} the approximate ILC reach for a $\sqrt{s}=0.5$ and
1 TeV machines by delineating the mass contours where $m_{\tw_1}$ and
$m_{\ttau_1}=250$ and 500~GeV (some additional region may be accessible
beyond this via $\tz_1\tz_2$ production \cite{nlc}). The bulk of the ILC
reach for a $\sqrt{s}=0.5$ TeV machine reaches to the $m_{1/2}\sim 500$
GeV level, corresponding to a gluino mass of $\sim 1150$~GeV.  A
$\sqrt{s}=1$~TeV ILC has a reach extending to $m_{1/2}\sim 1150$~GeV,
corresponding to a value of $m_{\tg}\sim 2600$~GeV. The enhanced reach
of ILC coupled to decreased reach of LHC in terms of $m_{1/2}$ means
that a 1~TeV ILC has a comparable reach to the LHC.

Within the HM2DM model, we have $M_1 \alt |\mu| \ll M_2$ at the weak
scale. Thus if gluinos ( remember that $m_{\tg} \sim 6M_1$) are
accessible at the LHC, it is reasonable to expect that $\tw_1$ as well
as $\tz_1$, $\tz_2$ and $\tz_3$ will be accessible at a TeV linear
collider.  The HM2DM framework will be readily distinguishable from the
mSUGRA model if chargino pair production is kinematically
accessible. Since the chargino is mainly a wino in mSUGRA, and a
higgsino in the HM2DM model, the total chargino pair production cross
section (for a given value of $m_{\tw_1}$) as well as
its dependence on the polarization of the
electron beam is very different. In the mSUGRA case, the expected 
cross section is significantly larger than in the HM2DM model, and further, 
drops to very low values as the electron beam is taken to be
increasingly right-handed, while this same dependence is comparably
milder for the higgsino-like $\tw_1$. This may be corroborated by studying
the polarization-dependence of neutralino pair production. Within mSUGRA,
the polarization dependence of the wino-like $\tz_2$ pairs is similar to
that of chargino pair production, while the higgsino-like $\tz_3$ and
$\tz_4$ are typically heavy. For the MHDM case realized in the HM2DM
framework, on the other hand, $\tz_1$, $\tz_2$ and $\tz_3$ are light and
mixed, so that not only are several $\tz_i\tz_j$ pairs accessible, the
polarization dependence and size of the cross sections is very
different.

This is illustrated in Fig.~\ref{fig:epol}, where 
we show various -ino pair production cross sections
accessible to a $\sqrt{s}=0.5$~TeV ILC versus beam polarization
$P_L(e^-)$ in the case of {\it a}) the mSUGRA model case and {\it b})
for the HM2DM1 case in Table~\ref{tab:m2dm}.  In frame {\it a}), we see
that only $\tw_1^+\tw_1^-$, $\tz_1\tz_2$ and $\tz_2\tz_2$ are accessible
to a $0.5$~TeV ILC, and that their cross sections precipitously drop
from readily observable values of tens or hundreds of femtobarns to
below 1~fb, as the electron beam polarization becomes increasingly
right-handed. In the HM2DM case shown in frame {\it b}), we see that as
anticipated many more -ino pair production reactions are accessible,
with vastly differing cross sections. The production of one chargino and
three neutralino states should be unambiguous. The relative size of the
various neutralino cross sections will be parameter-dependent: in the
present case the small size of $\sigma(\tz_2\tz_2)$ is a reflection of
the fact that the magnitudes of the ${\tilde h}_u$ and ${\tilde h}_d$
components of $\tz_2$ are nearly equal. The polarization dependence of
the (higgsino-like) neutralino production cross sections is also different
from frame {\it a}). A detailed study of the -ino
production reactions should allow the determination of $M_1$ and
$\mu$\cite{zerwas}.

%
\FIGURE[tbh]{
\epsfig{file=ee500-sigma-msug.eps,width=7cm} \hspace{0.5cm}
\epsfig{file=ee500-sigma-hm2dm1.eps,width=7cm} 
\caption{\label{fig:epol}
Cross section for $\tw_1^+\tw_1^-$ and $\tz_i\tz_j$
production at a $\sqrt{s}=0.5$ TeV
linear collider versus beam polarization parameter $P_L(e^-)$
for $m_0=300$~GeV, $m_{1/2}=300$~GeV, $A_0=0$, $\tan\beta =10$
and $\mu >0$ in the {\it a}) mSUGRA model and {\it b}) HM2DM
model with $M_2=900$~GeV.
}}

Another feature of the HM2DM model is that, since $M_2$ is large
and feeds into sfermion masses via the RG running from a universal
scalar mass, we expect
the lightest sfermions to be dominantly right-type, and the
heaviest are dominantly left-type. 
Within mSUGRA, the sfermions of the first two generations are
approximately degenerate, while  for the third generation
only the $\tau$-sleptons and top squarks are dominantly right-handed, 
while large top quark Yukawa effects make $\tb_1$ mostly the
left-type, as illustrated in Fig.~\ref{fig:mix}. 
We illustrate the beam polarization dependence of third generation
sfermion pair production
cross sections at a hypothetical $\sqrt{s}=1.5$ TeV ILC 
in Fig.~\ref{fig:ilc_sf} for {\it a}) the mSUGRA model
point in Table~\ref{tab:m2dm} and {\it b}) the case HM2DM2
case  in Table~\ref{tab:m2dm}. It is clear in the mSUGRA model
from the polarization dependence that in fact $\ttau_1$
is dominantly right-type, $\tb_1$ is dominantly left-type and
$\tst_1$ is mixed left-right. While the polarization dependence
for stau and stop pair production is qualitatively similar in the two
models, that for $b$-squark pair production is markedly different,
providing a clear signature for the qualitatively different value of $\theta_b$
in the HM2DM model.  
\FIGURE[tbh]{
\epsfig{file=eepol-sigma-msug_log.eps,width=7cm} \hspace{0.5cm} 
\epsfig{file=eepol-sigma-hm2dm_log.eps,width=7cm} 
\caption{\label{fig:ilc_sf}
Cross section for $\tst_1\bar{\tst}_1$, $\tb_1\bar{\tb}_1$
and $\ttau_1^+\ttau_1^-$ production at a $\sqrt{s}=1500$~GeV
linear collider versus beam polarization parameter $P_L(e^-)$
for $m_0=300$~GeV, $m_{1/2}=300$~GeV, $A_0=0$, $\tan\beta =10$
and $\mu >0$ in the {\it a}) mSUGRA model and {\it b}) HM2DM
model case  with $M_2=900$~GeV in Table~\ref{tab:m2dm}.
}}

\section{Summary and concluding remarks}\label{sec:conclude}

While the recently measured value of the CDM relic density can be
accommodated within the paradigm mSUGRA framework, much of the allowed
region lies at the edge of the parameter space as exemplified by
many studies in the $m_0-m_{1/2}$ plane. There are, however, several
one-parameter extensions of the mSUGRA model where it is possible to get
the observed value of $\Omega_{\rm CDM} h^2$ all over this plane for
essentially all values of $\tan\beta$ and $A_0$. These extensions
involve either the adjustment of the mass spectrum (so that the
neutralino annihilation rate is enhanced via an $s$-channel $A/H$
resonance or via co-annihilation with a charged sparticle) or the
adjustment of the neutralino composition (to obtain either MWDM
or MHDM). MHDM requires a reduced value of
$\mu^2$, and several models that lead to small $|\mu|$ have been
proposed. In this paper, we have pointed out a new mechanism for
obtaining the observed relic density: non-universal boundary
conditions with a large GUT scale value of the $SU(2)$ gaugino mass
parameter $|M_2|$ can lead to MHDM. We have also studied the broad
phenomenological implications of this scenario. 

Common to all scenarios with MHDM, our scenario has an -ino spectrum
where the lightest neutralino is a mixed bino-higgsino state, the
lighter chargino and the next two heavier neutralinos have large
higgsino content, while the heaviest chargino and neutralino are
dominantly wino-like with a mass $|M_2|$ considerably larger than the
other -ino masses. In the HM2DM model, $|M_2|$ is raised even further
and the wino states become very heavy and would likely only be 
produced at the LHC via cascade decays of $\tq_L$. 
The feature that distinguishes the
HM2DM model from other models with MHDM  is that {\it weak interaction
  effects} make the left-sfermions significantly heavier than their
right-siblings. This effect leads to a qualitative change in the
intrageneration mixing pattern of $b$-squarks: in the HM2DM model,
$\tb_1$ is dominantly right-handed, while in most models (with universal
mass parameters for $\tb_L$ and $\tb_R$) top-quark Yukawa coupling
effects make $\tb_1$ mostly left-handed. 

As in all models with MHDM, direct search experiments provide a
promising avenue once they reach a sensitivity to probe
neutralino-nucleon scattering cross sections $\alt 10^{-8}$~pb. With an
order of magnitude increase in sensitivity (projected at superCDMS or at
$\agt 100$~kg noble liquid detectors), there should be an observable
signal over most of the parameter space, while ton-sized noble element
detectors should be able to probe the entire parameter space. Our
conclusions for indirect searches are less definitive since, except for
IceCube type detectors, the signals depend on assumptions of the distribution
of the DM in our galactic halo. While IceCube may well be sensitive to
the signal, the best prospects for anti-particle detection appear to
come from the search for anti-deuterons. For the case that we examined,
the Pamela satellite that is currently gathering data may just be
sensitive enough to the signal from anti-protons for a favourable halo
profile, but less so for the positron signal. The dependence of the
gamma ray signal from our galactic center on the halo profile is too
large to draw strong conclusions about the observability in GLAST, but
perhaps a signal in anti-particle searches may make the situation
clearer.

We have discussed collider signals in Sec.~\ref{sec:col}. Prospect for a
discovery at the Tevatron are not encouraging within this
scenario. While the reach of the LHC is slightly degraded relative to
its reach in the mSUGRA model, experiments at the LHC should be
sensitive to gluino masses as large as 2750~GeV (2350~GeV) for a small
(moderate) value of $m_0$, assuming an integrated
luminosity of 100~fb. The TeV linear collider should be sensitive to a
discovery in a parameter space region very similar to the LHC. Searches
in the multi-jet plus opposite sign dileptons plus $\eslt$ channel will
be especially interesting as the dilepton mass distribution for same
flavour lepton pairs may not only allow the construction of more than
one mass edge (strongly suggestive of MHDM), but may also, via its shape,
provide indirect evidence for a higgsino-like $\tz_2$/$\tz_3$. 

It should be possible, at least in principle, to distinguish the HM2DM
model discussed here from other scenarios that also lead to agreement
with the observed relic density. Within a supersymmetric interpretation
of an observed signal at the LHC, the observation of more than one
dilepton mass edge would point to the existence of $\tz_2$ and $\tz_3$
with a relatively small mass splitting ($\le M_Z$) between $\tz_1$ and
these neutralinos. In turn, this would point to a small value of
$|\mu|$.\footnote{It is, of course, logically possible that $M_2$ is
just a bit smaller than $|\mu|$ so that all four neutralinos are
strongly mixed and accessible, but with $m_{\tz_4}-m_{\tz_1} > M_Z$. In
this case it would not be unreasonable to expect additional mass edges
from decays of the heavier neutralinos to $\tz_2$, or perhaps even to
$\tz_3$.}  If we further {\it assume} that the CDM density is {\it
saturated} by the LSP, we know that $\tz_1$ cannot be the higgsino.
Recall that for $M_1 \ll |\mu| \ll M_2$, $m_{\tz_1} \simeq M_1$ while
$m_{\tz_2}$ and $m_{\tz_3}$ would be on either side of $|\mu|$ so that
the value of $|\mu|-m_{\tz_1}$ must lie between the
$m_{\tz_3}-m_{\tz_1}$ and $m_{\tz_2}-m_{\tz_1}$ mass edges. We note,
however, that while the second inequality always holds within the HM2DM
model, $|\mu|$ may not always be that much larger than $M_1$ as, for
instance, in the HM2DM1 case. The mass edges nevertheless serve as a
semi-quantitative indicator of $|\mu|$ relative to $m_{\tz_1}$, and so
may provide corroborative evidence for consistency with the MHDM
hypothesis. A striking confirmation of the MHDM hypothesis could come
from the observation of a large cross section in direct DM search
experiments. While the dilepton mass edges may serve to separate out the
MHDM scenarios from those with MWDM or BWCA, they do not provide
evidence for the HM2DM framework, since this would likely require a
complete reconstruction of the -ino sector with a precision difficult at
the LHC.
Aside
from a determination of $M_2$ via that of $m_{\tw_2}$ (or $m_{\tz_4}$),
the smoking gun for the HM2DM framework would be observing a 
wino-like $\tw_2 /\tz_4$ and/or a large
splitting between the left and right sfermions, or determining that the
lighter $b$-squark state is mainly $\tb_R$. None of these appear
straightforward at the LHC. It is unlikely the heavier slepton or
chargino will be accessible at the LHC in this scenario, and squark mass
splittings and mixing angles are difficult to measure there. A high
energy electron-positron collider offers the best prospects for these
measurements, assuming, of course, that these sparticles are
kinematically accessible.

\acknowledgments

This research was supported in part by grants from the U.S. Department
of Energy. HB would like to thank the CERN theory group 
and LHC/cosmology visitor program for 
hospitality while this work was completed.

%

\end{document}